\documentclass[aps, prd, showpacs, nofootinbib, preprintnumbers,floatfix,letterpaper,longbibliography, reprint]{revtex4-1}
\usepackage[utf8]{inputenc}
\usepackage[colorlinks=true, linkcolor=blue, citecolor=teal]{hyperref}
\usepackage{bm}
\usepackage{latexsym,amssymb,amsmath,amsfonts}
\usepackage{graphicx}
\usepackage{longtable}
\usepackage{physics}
\usepackage{bm}
\usepackage{footnotebackref}
\usepackage{multirow}
\usepackage[usenames]{xcolor}
\usepackage{soul}
\usepackage{makecell}
\usepackage{graphicx}
\usepackage{amssymb}
\usepackage{amsmath}
\usepackage{longtable}
\usepackage{rotating}
\usepackage{color}
\usepackage{fancyhdr}
\usepackage{ifthen}
\usepackage{slashed}
\usepackage{xspace}
\usepackage{tabularx}
\usepackage{float}

\def\OGW{\Omega_\text{GW}}
\def \hpx{{\tt HEALPix} }
\def \hatOm {{\vb{\hat{\Omega}}}}

\begin{document}

\title{Jointly setting upper limits on multiple components of an anisotropic stochastic gravitational-wave background}

\author{Jishnu Suresh}
\email{jishnu@icrr.u-tokyo.ac.jp}
\affiliation{Institute for Cosmic Ray Research (ICRR), The University of Tokyo,
Kashiwa City, Chiba 277-8582, Japan} 

\author{Deepali Agarwal}
\email{deepali@iucaa.in}
\affiliation{Inter-University Centre for Astronomy and Astrophysics (IUCAA), Pune 411007, India}

\author{Sanjit Mitra}
\email{sanjit@iucaa.in}
\affiliation{Inter-University Centre for Astronomy and Astrophysics (IUCAA), Pune 411007, India}
\begin{abstract}
With the increasing sensitivities of the gravitational wave (GW) detectors and more detectors joining the international network, the chances of detection of a stochastic GW background (SGWB) are progressively increasing. Different astrophysical and cosmological processes are likely to give rise to backgrounds with distinct spectral signatures and distributions on the sky. The observed background will therefore be a superposition of these components. Hence, one of the first questions that will come up after the first detection of a SGWB will likely be about identifying the dominant components and their distributions on the sky. Both these questions were addressed separately in the literature, namely, how to separate components of isotropic backgrounds and how to probe the anisotropy of a single component. Here, we address the question of how to separate distinct anisotropic backgrounds with (sufficiently) different spectral shapes. We first obtain the combined Fisher information matrix from folded data using an efficient analysis pipeline {\tt{PyStoch}}, which incorporates covariances between pixels and spectral indices. This is necessary for estimating the detection statistic and setting upper limits. However, based on a recent study, we ignore the pixel-to-pixel noise covariance that does not have a significant effect on the results at the present sensitivity levels of the detectors. We establish the validity of our formalism using injection studies. We show that the joint analysis accurately separates and estimates backgrounds with different spectral shapes and different sky distributions with no major bias. This does come at the cost of increased variance. Thus making the joint upper limits safer, though less strict than the individual analysis. We finally set joint upper limits on the multicomponent anisotropic background using Advanced LIGO data taken up to the first half of the third observing run.
\end{abstract}

\maketitle
\section{Introduction}
A stochastic gravitational-wave background (SGWB)~\cite{AllenOttewill,Romano2017}, created by the superposition of signals from unresolved astrophysical and cosmological sources~\cite{2018PhRvL.120i1101A,2011PhRvD..84h4004R, 2011PhRvD..84l4037M, 2011ApJ73986Z, 2012PhRvD..85j4024W, 2013MNRAS.431..882Z,2016PhRvD..94j3011D, 2021PhRvD.103d3002P,hotspot,HUGHES201486,2018arXiv180710620C,PhysRevD.88.062005,PhysRevD.50.1157,PhysRevD.85.023534,CROWDER201366}, are one among the most interesting targets for the current and upcoming gravitational wave detectors. Due to the nonuniform distribution of these astrophysical sources in the local Universe, the background can be significantly anisotropic~\cite{PhysRevD.98.063509,2014PhRvD..89h4076M, 2018PhRvD..98f3501J,2012PhRvD..86j4007R, 2013PhRvD..87d2002W,2013PhRvD..87f3004L,PhysRevD.96.103019}. Detection of such a SGWB seems promising in the near future and developing algorithms to probe the anisotropy in the background is, therefore, a highly relevant topic of research. Each source, generated by different phenomena, contributes to the anisotropic SGWB differently. Contributions from the individual components can be characterized by specific spectral shapes~\cite{CowardTania}. Upper limits have been placed on the anisotropic SGWB for primarily three spectral indices from the past observing runs of Advanced LIGO and Advanced Virgo~\cite{O1paper,O2paper,o3_direction,Renzini_aniso}. 

These directional upper limits are placed for one spectral index at a time, neglecting the existence of the other components and thereby the covariance between different spectral components. The presence of one source contribution, however, can affect the estimation of the other components. Consider a situation where we filter the GW data for multiple SGWB components separately. We will end up overestimating the amplitudes of the individual components and underestimating the error bars. Hence, for the possibility of coexistence of multiple spectral components, one must estimate the amplitudes of individual components jointly, a joint-index multicomponent estimation is therefore necessary. This need does not supersede the existing results for single indices if one component is much stronger than the others. However, for scenarios where the strength of at least one component is not negligible, a joint estimation will be necessary. For instance, if we want to set an upper limit on the average eccentricity of extragalactic pulsars by observing the stochastic background created by them and, since we know that the same galaxy clusters also host the binary mergers (though the proportions may vary across the clusters), the latter cannot be ignored while setting the former upper limit~\cite{Talukder_NS}. On the other hand, one can also get limits on an unmodeled spectrum~\cite{Renzini_aniso,pystoch} by estimating the background in multiple independent frequency bins. This may not separate different signals but could reduce bias in the recovery that one might see if one uses a single spectrum over the whole analysis, though the variance in each bin will be significantly higher than that in the full broadband analysis.

Methods have been proposed in the past to disentangle these components~\cite{Ungarelli,vukPE,Katarina_2021,Sylvia}. These methods are appropriate for estimating the parameters associated with different SGWB models. However, with the increase in SGWB components, we have to deal with likelihood functions in multidimensional parameter space. In such situations, techniques like Markov-chain Monte Carlo or nested sampling will be needed to estimate the model parameters. Recently, a new method has been proposed in \citet{Parida_isotropic}, where the Maximum-Likelihood estimation problem for a multicomponent isotropic background was transformed to a linear deconvolution problem. This makes the estimation of the SGWB components trivial and computationally cheap. This novel component separation method was extended to the anisotropic SGWB case~\cite{Parida_anisotropic}, where one can jointly estimate the amplitudes of the many components and set a directional upper limit. In this present work, we make use of the folded data~\cite{fold} and the efficient {\tt{PyStoch}}~\cite{pystoch,pystochSpH} pipeline to produce the first simultaneous $95\%$ confidence level directional upper limits on multicomponent anisotropic SGWB from the Advanced LIGO's observational data taken up to the first half of the third observing run.

The paper is organized as follows: In Section~\ref{sec:method} we discuss the component separation method. Sec.~\ref{sec:data} briefly summarizes the dataset used. We describe the injection study performed to test the algorithm in Sec.~\ref{sec:injection}. We also run the component separation algorithm on Advanced LIGO's observational data taken up to the first half of the third observing run. The details on this run along with the conclusions are described in Sec.~\ref{sec:result}

\section{Methods}
\label{sec:method}
Assuming the SGWB to be unpolarized, Gaussian and stationary, the quadratic expectation value of the GW strain distribution $h_A(f,\hatOm)$ across different sky directions and frequencies can be written as
\begin{equation}
    \langle h_A^*(f, \hatOm) \, h_{A'}(f', \hatOm') \rangle = \frac{1}{4} \mathcal{P}(f, \hatOm) \, \delta_{AA'} \, \delta(f-f') \delta(\hatOm, \hatOm') \, ,
\end{equation}
where $\mathcal{P}(f, \hatOm)$ denotes the true anisotropy in the sky for a polarization $A$ (from here on-wards, we assume that both polarizations contribute equally). Now we can write the SGWB energy density having the units of the dimensionless energy density parameter per steradian as
\begin{equation}
\OGW (f,\hatOm)=\frac{2\,\pi^2}{3H_0^2}\,f^3\,\mathcal{P}(f,\hatOm)\,.
\end{equation}
It is also useful to write the above equation in terms of energy flux, which has units of ${\rm erg \, cm^{-2} \, s^{-1} \, Hz^{-1}}$ as
\begin{equation}
\mathcal{F} (f,\hatOm)=\frac{c^3\,\pi}{4 G}\,f^2\,\mathcal{P}(f,\hatOm)\,.
\end{equation}
 Assuming that the source frequency power spectrum remains constant, and separable from its angular dependency, we can write the true sky as
\begin{equation}
 \mathcal{P}(f, \hatOm) = H(f) \, \mathcal{P}(\hatOm)\, ,
\end{equation}
where $H(f)$ is the spectral shape of the background. If we assume that the SGWB is constituted by multiple components, denoted by the spectral index $\alpha$, with a known spectral shape (we assume power-law spectrum throughout this work), 
\begin{equation}
    H_{\alpha} (f) = \left(\frac{f}{f_{\rm ref}}\right)^{\alpha -3},
\end{equation}
here $f_{\rm ref}$ is a reference frequency at which the detector is most sensitive is set as 25 Hz. Then the amplitude of the SGWB intensity can be rewritten as
\begin{equation}
 \mathcal{P}(f, \hatOm) = \sum_{\alpha} H_{\alpha} (f) \, \mathcal{P}^{\alpha}(\hatOm)\, .
 \label{eq:multi_comp}
\end{equation}
This expression differs from the usual form, $\mathcal{P}(f, \hatOm) = H (f) \, \mathcal{P}(\hatOm)$ used in GW radiometer algebra~\cite{fold,Mitra07,ballmer06,fold}. The corresponding changes in the algebra are given below.

We start with the usual cross-spectral density (CSD), $\mathcal{C}^\mathcal{I} (t;f) := \widetilde{s}_{\mathcal{I}_1}^*(t;f) \widetilde{s}_{\mathcal{I}_2}(t;f)$, of data from baseline $\mathcal{I}$ formed by the pair of detectors $\mathcal{I}_1$ and $\mathcal{I}_2$, where $\widetilde{s}_{\mathcal{I}_{1,2}}$ are the short-term Fourier transforms (SFTs) of time-series data obtained from the data segments of duration $\tau$ at time $t$. The usual expectation of the CSD~\cite{fold} then gets modified to,
\begin{equation}
    \langle \mathcal{C}^\mathcal{I} \rangle = \tau \sum_{\alpha} H_{\alpha} (f) \, \gamma_{ft,u}^{I} \, \mathcal{P}_u^\alpha \,,
\end{equation}
where the sky-maps have been expanded in the basis $e_u(\mathbf{\hat \Omega})$ as $\mathcal{P}^\alpha(\hatOm) = \sum_u \mathcal{P}^\alpha_u e_u(\mathbf{\hat \Omega})$ and $\gamma_{ft,u}^{I}$ is the generalized overlap reduction function (ORF)~\cite{christ92,ORF_Finn}, that accounts for the mismatch between the response functions of the detectors and the delay in signal arrival times, defined as,
\begin{equation}
    \gamma_{ft,u}^{I} \ := \ \sum_{A} \int d \mathbf{\hat \Omega} F^{A}_{\mathcal{I}_1}(\mathbf{\hat \Omega},t) F^{A}_{\mathcal{I}_2}(\mathbf{\hat\Omega},t) \, e^{2\pi i f \frac{\mathbf{\hat \Omega}\cdot {\mathbf{\Delta x}_I (t)}}{c}} \, e_u(\mathbf{\hat \Omega}) \,.
\end{equation}
In the above equation the polarizations are denoted by $A=+,\times$, and $F^{A}_{\mathcal{I}_{1,2}}(\mathbf{\hat \Omega},t)$ denotes the respective antenna pattern functions. Here $\mathbf{\Delta x}_I (t)$ is the separation vector between the detectors. In pixel basis, the ORF takes the form,
\begin{equation}
    \gamma_{ft,u}^{I} \ = \ \sum_{A} F^{A}_{\mathcal{I}_1}(\mathbf{\hat \Omega}_u,t) F^{A}_{\mathcal{I}_2}(\mathbf{\hat\Omega}_u,t) \, e^{2\pi i f \frac{\mathbf{\hat \Omega}_u\cdot {\mathbf{\Delta x}_I (t)}}{c}} \,,
\end{equation}
where $\mathbf{\hat \Omega}_u$ is the direction of pixel $u$.
The folded CSDs~\cite{fold} average the SFTs over many sidereal days and can be assumed to be Gaussian.\footnote{The same argument can be extended to show that even unfolded CSDs would lead to the same Gaussian Likelihood function, as long as the observation period is tens of days or longer.} Hence, for a signal model that allows for a superposition of multiple GWB components [as in Eq.~(\ref{eq:multi_comp})], and assuming that the detector noise spectra are well estimated, the likelihood function can be written as
\begin{equation}
     p(\mathcal{C}^\mathcal{I}|\, \mathcal{P}^\alpha_u) \propto   \, {\rm exp} \Big[-\frac{1}{2} (\mathcal{C}^\mathcal{I} - \langle \mathcal{C}^\mathcal{I} \rangle)^*  \Big.  \Big. \mathcal{N}^{-1}  (\mathcal{C}^\mathcal{I} - \langle \mathcal{C}^\mathcal{I} \rangle) \Big] \, ,
\end{equation}
where the covariance of $\mathcal{C}^\mathcal{I}$ for different times and frequencies is given by the matrix $\mathcal{N}$. This covariance matrix turns out to be diagonal, that is, data segments from different times and frequencies are uncorrelated. Data from different baselines $\mathcal{I}$ are also uncorrelated~\cite{allen-romano,S5_Dir}.

Then following the standard maximum-likelihood (ML) method for mapping the GW intensity~\cite{fold,Mitra07,Eric09} in a general basis, the clean map is given by
\begin{equation}
 \mathcal{\hat{P}}_{u}^\alpha \ \equiv \ \hat{\vb{\bm{\mathcal{P}}}} \ = \ \vb{\Gamma}^{-1} \cdot \vb{X}
\label{eq:Clean_map}
\end{equation}
where, the ``dirty map,''
\begin{equation}
\vb{X} \equiv X^{\alpha} _u =  \sum_{Ift} \gamma^{I*}_{ft,u} \frac{ H_{\alpha}(f)} {P_{\mathcal{I}_1}(t;f) P_{\mathcal{I}_2}(t;f)}  \mathcal{C}^{I} (t;f)\,,
\label{eq:Dirty_map}
\end{equation}
and the Fisher information matrix,
\begin{equation}
\vb{\Gamma} \equiv \Gamma^{\alpha \beta} _{uu'} = \sum_{Ift} \frac{H_{\alpha}(f) H_{\beta}(f)}{P_{\mathcal{I}_1}(t;f) \, P_{\mathcal{I}_2}(t;f)} \,\gamma^{I*}_{ft,u} \, \gamma^{I}_{ft,u'} \, .
\label{eq:Fisher}
\end{equation}
In the above equations $P_{\mathcal{I}_{1,2}}(t;f)$ are the one-sided noise power spectral density of the individual detectors. From Eq.~(\ref{eq:Fisher}), it is evident that the Fisher matrix is proportional to the noise-weighted inner product of the spectral shapes $H_{\alpha}(f)$ and $H_{\beta}(f)$. The clean map estimator takes the simple form given in Eq.~(\ref{eq:Clean_map}), when the spectral shapes are not degenerate. However
direct inversion of the Fisher matrix could lead to numerical errors. This may demand some form of conditioning to perform the matrix inversion~\cite{Eric09,regDeconv,O1O2Folded}.

\section{Data}
\label{sec:data}
To perform component separation and to compute the upper limits using the joint-index formalism, we use the strain data~\footnote{The data that support the findings of this study are openly available at https://www.gw-openscience.org/data/} from the first (O1), second (O2), and first-half of third (O3a) observing runs of Advanced LIGO detectors located in Hanford (H) and Livingston (L)~\cite{O1O2openData}. The fetched data have been processed and conditioned in the same way as was done in the previous analyses~\cite{S5_Dir,O1paper,O2paper,o3_direction} to obtain the cross-spectral density and the corresponding variance, with a coarse-grained resolution of 1/32 Hz. Following the steps described in~\citet{fold}, these datasets are then folded into one sidereal day. We then perform the radiometer analysis ``blindly" on the folded dataset to obtain the dirty map and the Fisher matrix corresponding to each observational run using {\tt{PyStoch}}. One can obtain the combined Fisher matrix and dirty map from these data by following~\cite{Romano2017},
\begin{eqnarray}
 \vb{\Gamma} &=& \vb{\Gamma}^{O1}+\vb{\Gamma}^{O2}+\vb{\Gamma}^{O3a}\,, \nonumber \\
 \vb{X} &=& \vb{X}^{O1} + \vb{X}^{O2} + \vb{X}^{O3a}  \,.
\label{eq:combined_dirty_fisher}
\end{eqnarray}

 \begin{figure}
    \centering
    \includegraphics[width =\columnwidth]{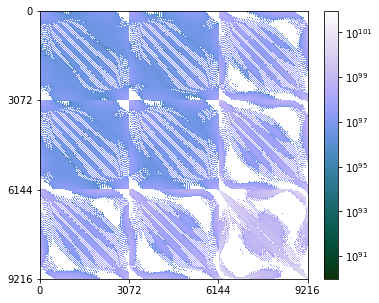}
    \caption{Coupling matrix showing the covariance between the spectral shapes. Each 3072x3072 block represent the Fisher matrix for one particular spectral index. The indices (0,2/3,3) are increasing from left to right and top to bottom. These are computed and combined following Eq.~(\ref{eq:matrix_convolution}).}
    \label{fig:Coupling_Mat}
\end{figure}

\begin{figure}[b]
\centering
\includegraphics[width = 0.23\textwidth]{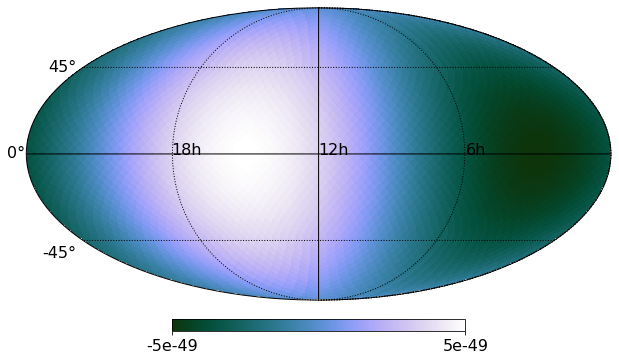}
\includegraphics[width = 0.23\textwidth]{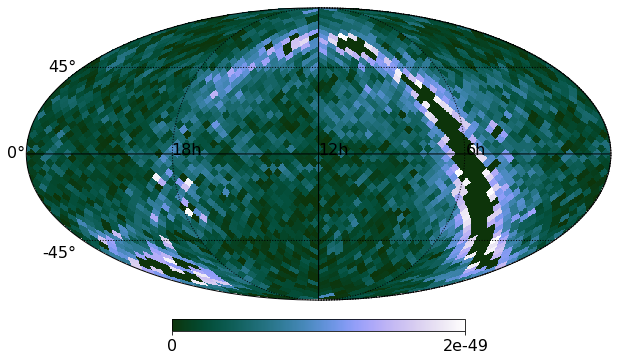}\\
\includegraphics[width = 0.23\textwidth]{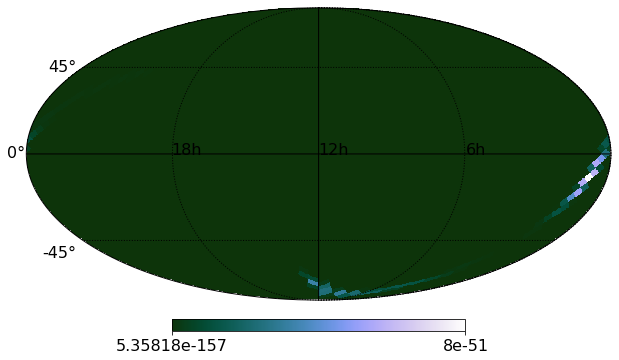}
\includegraphics[width = 0.23\textwidth]{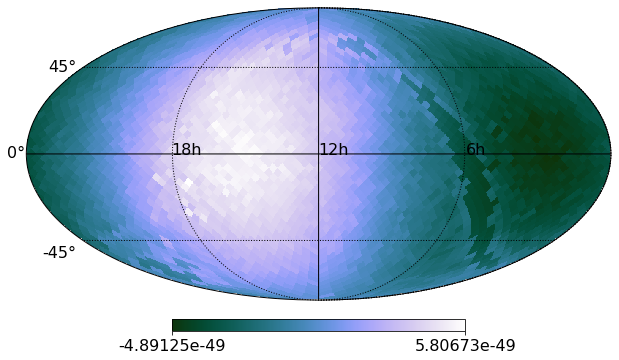}
\caption{Input maps considered for the injection study. From left, sky maps represent the source maps for $\alpha = 0, 2/3, 3$ and the combined input maps. All the maps are represented as a color bar plot on a Mollweide projection of the sky in ecliptic coordinates.}
\label{fig:injection}
\end{figure}

\section{Injection study}
\label{sec:injection}
We now demonstrate the component separation method for an anisotropic background using injections. In all the previous stochastic analyses performed on the LIGO's observing runs, three values of $\alpha$ have been considered. These spectral indices are related to different SGWB models: $\alpha=0$, corresponding to different cosmological models~\cite{Caprini2018}; $\alpha=2/3$ is related to an astrophysical background which is dominated by compact binary coalescence~\cite{sesana2008}; and $\alpha=3$, indicating a flat strain spectrum~\cite{gw150914_stoch} which stand for sources like millisecond pulsars and magnetars. Throughout this work, we will consider only these three spectral dependencies to estimate the SGWB amplitudes. 

Now for the three spectral index case, the clean map [Eq.~(\ref{eq:Clean_map})] can be obtained by solving the convolution equation in the matrix form
\begin{equation}
\begin{bmatrix}
    X^{\alpha1}_{u}  \\
    X^{\alpha2}_{u}  \\
    X^{\alpha3}_{u} 
\end{bmatrix}
=
\begin{bmatrix}
    \Gamma^{\alpha1 \alpha1}_{uu'} & \Gamma^{\alpha1 \alpha2}_{uu'} & \Gamma^{\alpha1 \alpha3}_{uu'}  \\
    \Gamma^{\alpha2 \alpha1}_{uu'} & \Gamma^{\alpha2 \alpha2}_{uu'} & \Gamma^{\alpha2 \alpha3}_{uu'}  \\
    \Gamma^{\alpha3 \alpha1}_{uu'} & \Gamma^{\alpha3 \alpha2}_{uu'} & \Gamma^{\alpha3 \alpha3}_{uu'} 
\end{bmatrix}
\begin{bmatrix}
   \mathcal{\hat{P}} ^{\alpha1}_{u'}  \\
   \mathcal{\hat{P}} ^{\alpha2}_{u'}  \\
   \mathcal{\hat{P}} ^{\alpha3}_{u'} \,
\end{bmatrix} \,.
\label{eq:matrix_convolution}
\end{equation}
The $3\times3$ matrix in the above equation denotes the coupling matrix $C^{\alpha \beta} _{uu'}$ formed by combining the Fisher information matrices for all possible combinations of the spectral indices. Here $\{\alpha,\beta\} \in \{\alpha1,\alpha2,\alpha3\}$ denotes the spectral index which assumes different values for different contributing sources (each elements in the above equation represents either a vector or tensor with dimension of $u$ and $u'$). We can now rewrite Eq.~(\ref{eq:Clean_map}), the ML solution of the convolution equation, in terms of the coupling matrix as
\begin{equation}
    \hat{\mathcal{P}} ^{\alpha} _{u} \ = \ \sum_{u'\beta} \left[ C ^{-1} \right] ^{\alpha \beta} _{uu'} \ X^{\beta} _{u'} \ \ .
    \label{eq:deconv_c}
\end{equation}
However, in reality, the coupling matrix $\vb{C}$ is sparse in nature due to the covariance between the spectral shapes (these effects could further boosted by the pixel-to-pixel covarinace). Hence the direct inversion of $\vb{C}$ will lead to numerical errors. We resolve this by invoking a \textit{preconditioning} of this coupling matrix. We achieve this by defining,
\begin{equation}
    \vb{C'} \equiv C'^{\alpha \beta} _{uu'} = \frac{\Gamma^{\alpha \beta}_{uu'}}{\sqrt{\Gamma^{\alpha \alpha}_{uu} \, \Gamma^{\beta \beta}_{u'u'}}} \,,
    \label{eq:c_prime}
\end{equation}
where the $\vb{C'}$ becomes a block matrix with a unit diagonal and the off-diagonal components are positive numbers. To show the preconditioned coupling matrix elements, we chose one particular direction (100th pixel in a \hpx resolution of $N_{\mbox{side}} = 16$ following the RING~\cite{HEALPix} ordering scheme.), and then for that particular pixel $\vb{C'}$ equates to
\begin{equation}
C'^{\alpha \beta} _{uu} = 
    \begin{bmatrix}
    1.0 & 0.96477635 & 0.32658296  \\
    0.96477635 & 1.0 & 0.47682514  \\
    0.32658296 & 0.47682514 & 1.0 
\end{bmatrix}\,.
\end{equation}
As it is evident from the above coupling matrix example, the $\alpha=0$ and $\alpha=2/3$ components are highly correlated. Now we can rewrite Eq.~(\ref{eq:deconv_c}), in terms of preconditioned coupling matrix as
\begin{equation}
    \hat{\mathcal{P}} ^{\alpha} _{u} \ = \ ~ \sum_{u' \beta}   \left[ C'^{-1} \right]^{{\alpha \beta}} _{uu'} \  \ \frac{X'^{\beta} _{u'}} {\sqrt{\Gamma^{\alpha \alpha}_{uu}}} \ \ ,
    \label{eq:deconv_c_prime-gen}
\end{equation}
where $X'^{\beta} _{u'} = X^{\beta} _{u'}/\sqrt{\Gamma^{\beta \beta}_{u'u'}}$.

It was recently shown~\cite{O1O2Folded} that at the current sensitivity level of the detectors, the contribution from the off-diagonal components of the pixel-to-pixel covariance matrix on the upper limits can be ignored. Following which we limit our analysis to retaining only the diagonal components of the pixel-to-pixel covariance matrix (however, as the detector sensitivity and number of detectors increase, we may have to incorporate these effects as well). Given we are neglecting the pixel-to-pixel covariance, Eq.~(\ref{eq:deconv_c_prime-gen}) takes a simple form for a given pixel as
\begin{equation}
    \hat{\mathcal{P}} ^{\alpha} _{u} \ = \  ~ \sum_\beta \left[ C'_{uu} \right]^{-1}_{{\alpha \beta}}  \ \frac{X'^{\beta} _{u'}} {\sqrt{\Gamma^{\alpha \alpha}_{uu}}} \ \ ,
    \label{eq:deconv_c_prime}
\end{equation}
We can now solve the above linear convolution equations [Eq.~(\ref{eq:deconv_c_prime})] with $\vb{C'}$ to get the estimate of the SGWB sky.

To demonstrate our method, we first consider the O1+O2+O3a data from the HL baseline. Figure~\ref{fig:Coupling_Mat} shows the coupling matrix for all the three spectral indices $(0,2/3,3)$ computed from this dataset. Now we consider source signals from different input maps to verify the joint-index multicomponent separation and hence the signal reconstruction. We use three types of toy models for injections: one containing a dipolelike sky pattern with $\alpha=0$; an extended source that mimics the diffuse part of the Milky Way Galaxy, whose spectral dependency is characterized by $\alpha=2/3$; and a sky-patch following $\alpha=3$ spectral shape. We work at a \hpx resolution $N_{\mbox{side}} = 16$ throughout the map-making process. Figure~\ref{fig:injection} shows the injected sky maps for all three cases along with the sky map obtained by combining these individual sources. The injection strengths we choose are such that the background signal is well above the noise level.

\begin{figure*}[t]
\centering
\includegraphics[width = 0.32\textwidth]{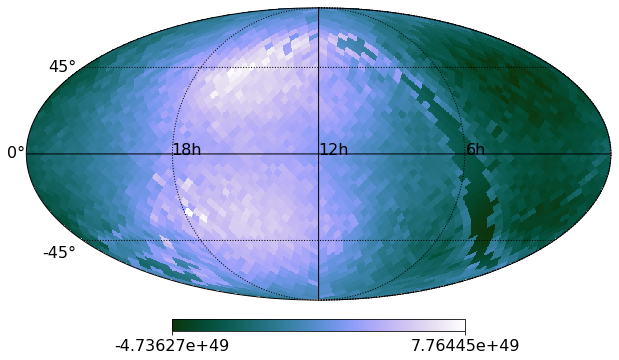}
\includegraphics[width = 0.32\textwidth]{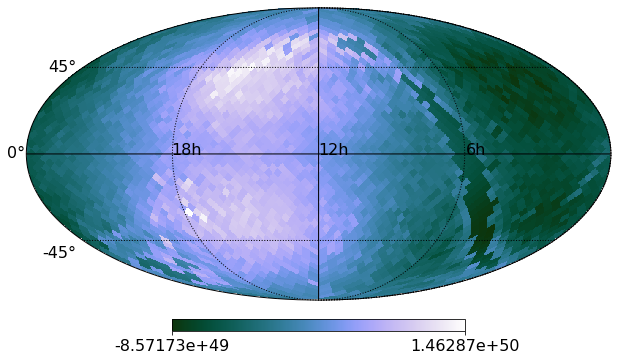} 
\includegraphics[width = 0.32\textwidth]{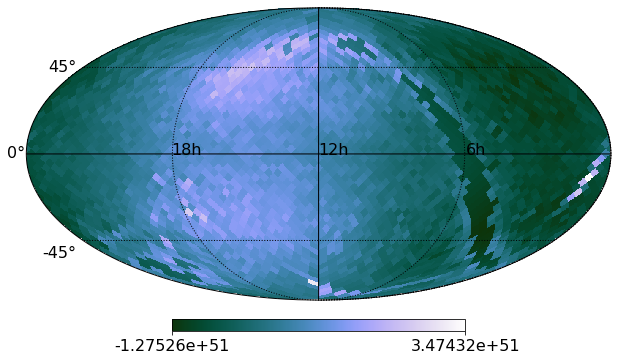} \\
\includegraphics[width = 0.32\textwidth]{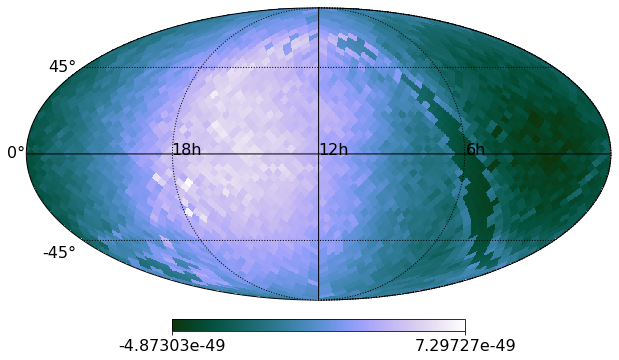}
\includegraphics[width = 0.32\textwidth]{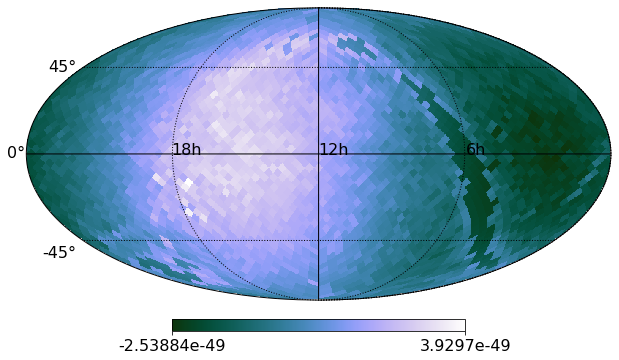} 
\includegraphics[width = 0.32\textwidth]{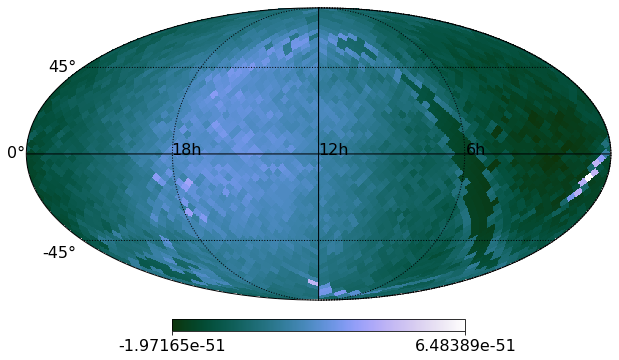} \\
\includegraphics[width = 0.32\textwidth]{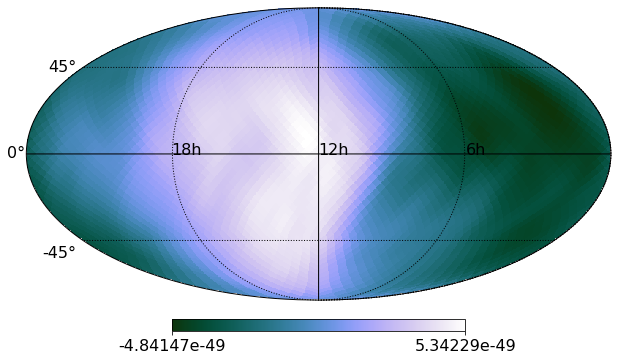}
\includegraphics[width = 0.32\textwidth]{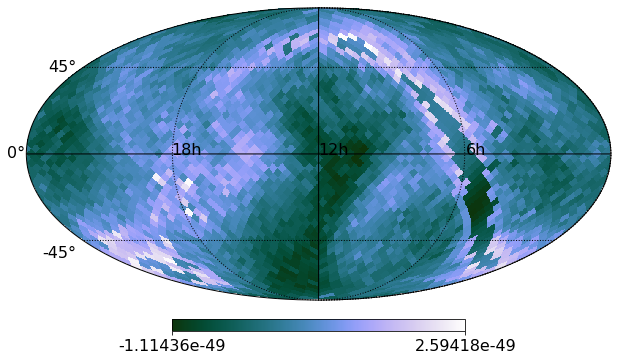} 
\includegraphics[width = 0.32\textwidth]{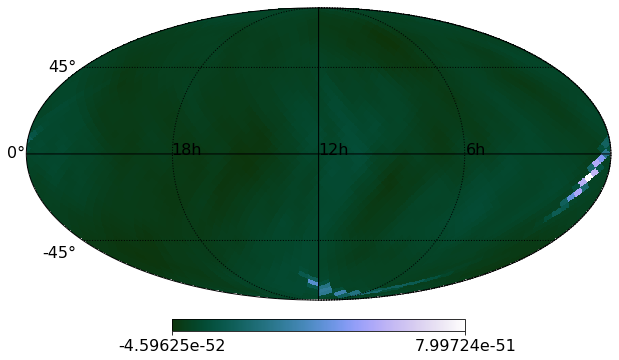}
\caption{Recovery of the injected source map using single index and multi-index estimations for $\alpha=0,2/3,3$ from left to right. The first row shows the dirty maps obtained by the optimal filtering assuming a spectral shape. The second row shows the recovered semiclean map with the conventional single-index component separation. The third row depicts the semiclean map from a joint-index multicomponent analysis. All the maps are represented as a color bar plot on a Mollweide projection of the sky in ecliptic coordinates.}
\label{fig:recovery_injection}
\end{figure*}

We now apply the usual radiometer map-making method~\cite{pystoch} to the folded dataset with injections. In the radiometer analysis, we apply a time-varying phase delay to perform earth-rotation synthesis imaging and assign weights to the frequencies proportional to the expected source spectrum and inverse proportional to the detector noise spectrum~\cite{allen-romano}, to obtain the dirty map. The dirty maps obtained for each choice of the spectral functions are shown in the first row of Fig.~\ref{fig:recovery_injection}. After the dirty map creation, following the typical LIGO-Virgo-KAGRA (LVK) collaboration stochastic analyses, we neglect the pixel-to-pixel correlations~\footnote{We may have to consider a regularized inversion~\cite{Eric09,regDeconv} of the coupling matrix if we want to take into account the pixel-to-pixel correlations and perhaps for data from more and/or sensitive detectors.} and obtain the \textit{semiclean map} from both the conventional (``single-index") and the new joint-index multicomponent (``joint-index") analysis.

The second row of Fig.~\ref{fig:recovery_injection} shows the semiclean SGWB sky maps for all the three spectral indices for the single-index results. Here we assume that each SGWB spectral components are independent. Comparing with the injected source strength (see Fig.~\ref{fig:injection}), it is apparent that we overestimate the background of each SGWB component for the single-index analysis. This is not surprising since we neglect the correlated background coming from other components when we choose a particular spectral function $H(f)$. Hence this severely biased method is not the correct way to extract the SGWB when the background components for the other spectral indices cannot be ignored.

We now use our proposed prescription and obtain the semiclean map by properly taking into account the covariance between different spectral shapes. For this, we solve the equation given in Eq.~(\ref{eq:deconv_c_prime}) by applying  preconditioning [Eq.~(\ref{eq:c_prime})] on the coupling matrix. For each pixel, we invert the $3\times3$ covariance matrix comparing different spectral indices. The third row of Fig.~\ref{fig:recovery_injection} shows thus obtained semiclean maps. It is evident from the maps that joint-index successfully undo the correlation between the three spectral shapes. To evaluate the accuracy of this recovery we have to use a certain quantitative measure, which can be challenging to define~\cite{Mitra07,regDeconv}.

\begin{table}[b]
    \centering
    \begin{tabular}{c|c|c}
        \hline\hline
        \multicolumn{3}{c}{Normalized Mean Squared Error ($\times 10^{-5}$)} \\
        \hline
         Spectral shape $\alpha$  & Single-index & Joint-index  \\
         \hline
          0 & 4.02 & 2.45 \\
         \hline
         2/3 & 230.9 & 21.88 \\
         \hline
          3 & 548.2 & 12.15 \\
         \hline\hline
    \end{tabular}
    \caption{Quality of the source reconstruction as a function of NMSE for the semiclean maps from single-index and joint-index component estimations.}
     \label{tab:nmse}
\end{table}
In this work we use the normalized mean squared error (NMSE) to quantify the deviation of the reconstructed semiclean map from the input map. If $\mathcal{A}$, $\mathcal{B}$ are respectively the input map and the reconstructed map, NMSE is defined as,
\begin{equation}
     \mbox{NMSE}=\frac{\norm{{\mathcal{A}}-{{\mathcal{B}}}}^2}{\norm{\mathcal{A}}^2} \,.
     \label{eq:NMSE}
\end{equation}
A better recovery is indicated by the value of the NMSE being closer to zero. Table~\ref{tab:nmse} summarizes and compares the quality of the reconstructed sky map as a function of NMSE. The values of NMSE for the joint-index multicomponent separation case being close to 0, corroborates the importance of covariance between different components. 

\begin{figure*}
\centering
\includegraphics[width = 0.32\textwidth]{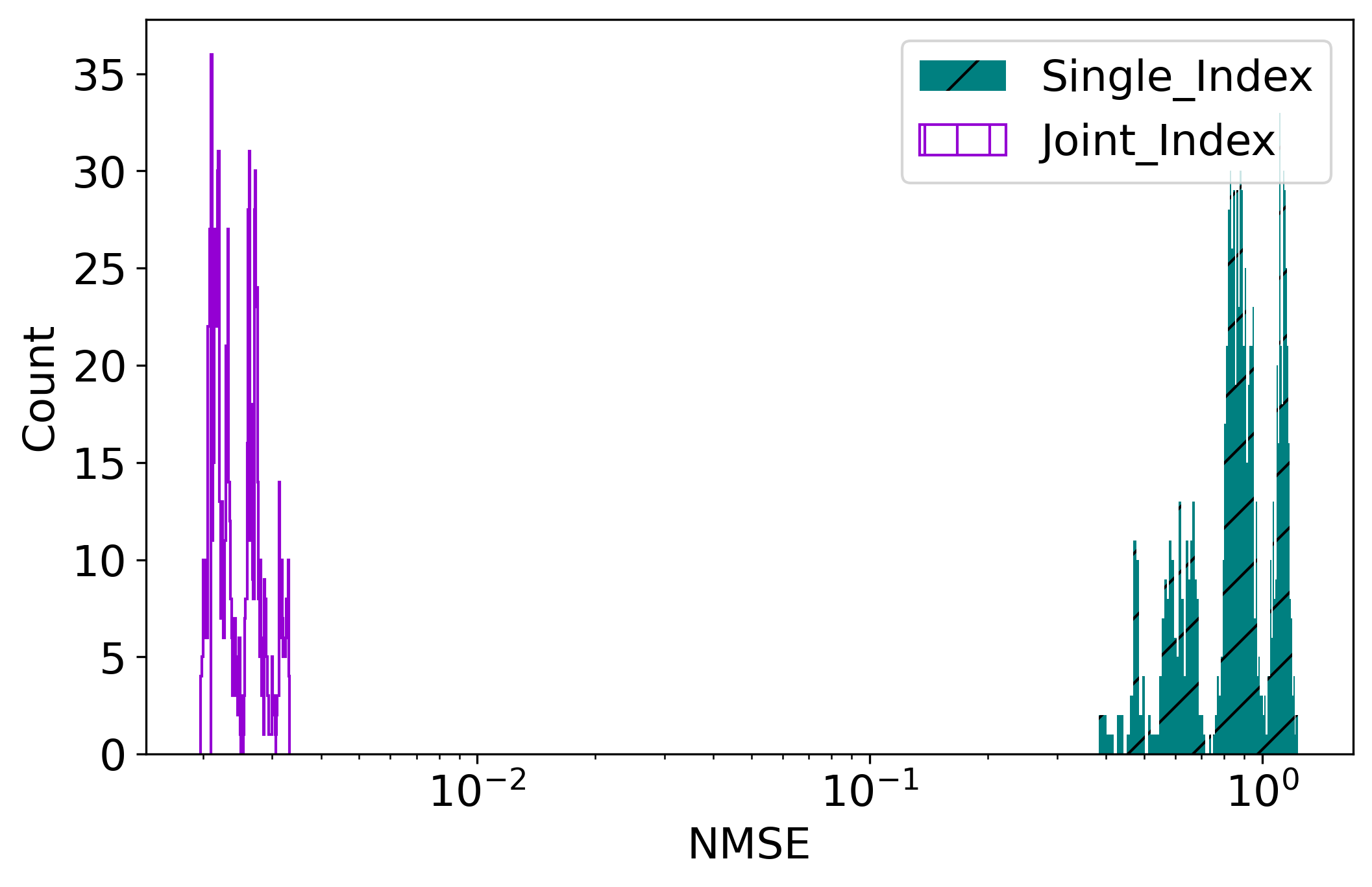}
\includegraphics[width = 0.32\textwidth]{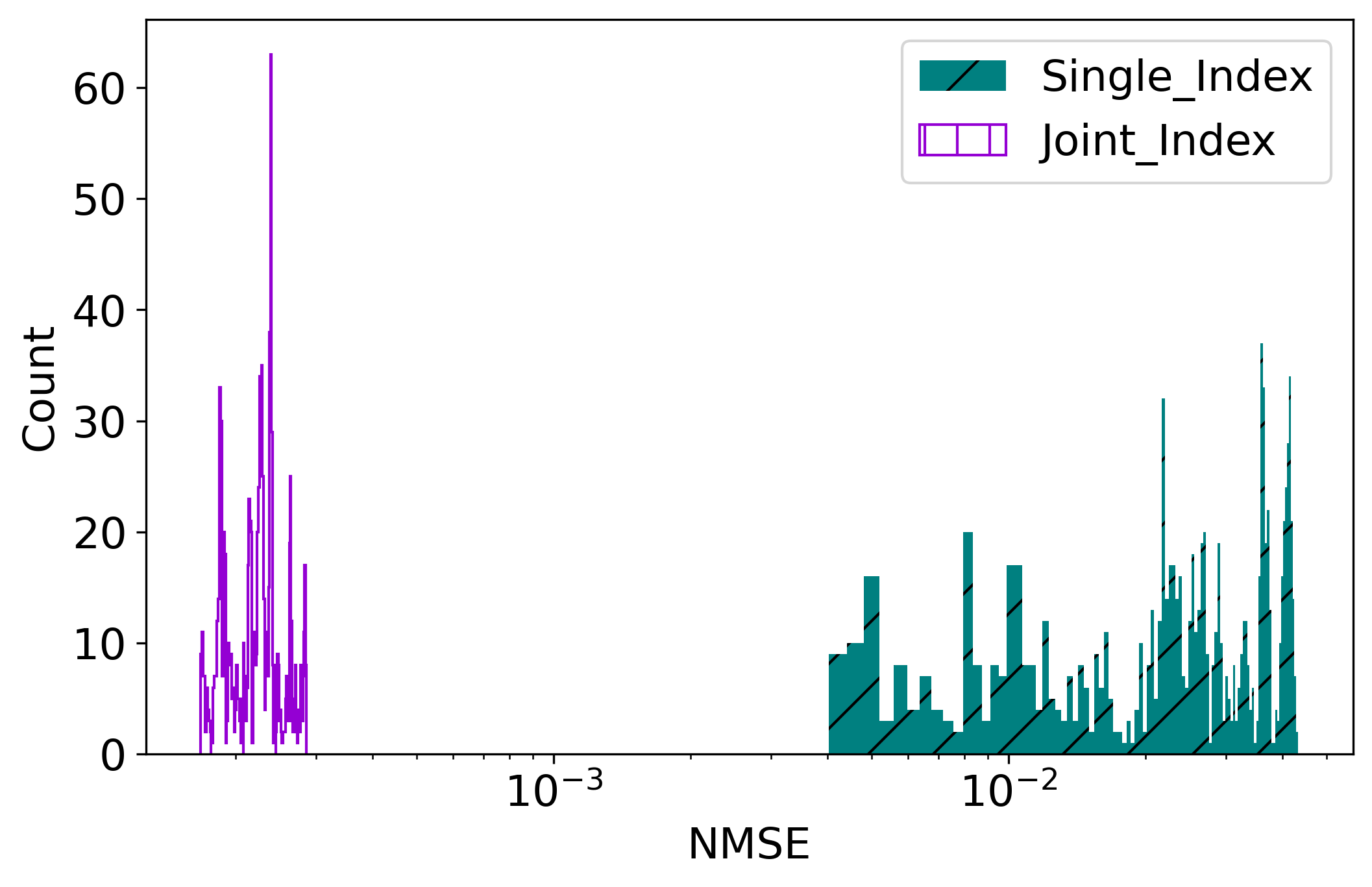}
\includegraphics[width = 0.32\textwidth]{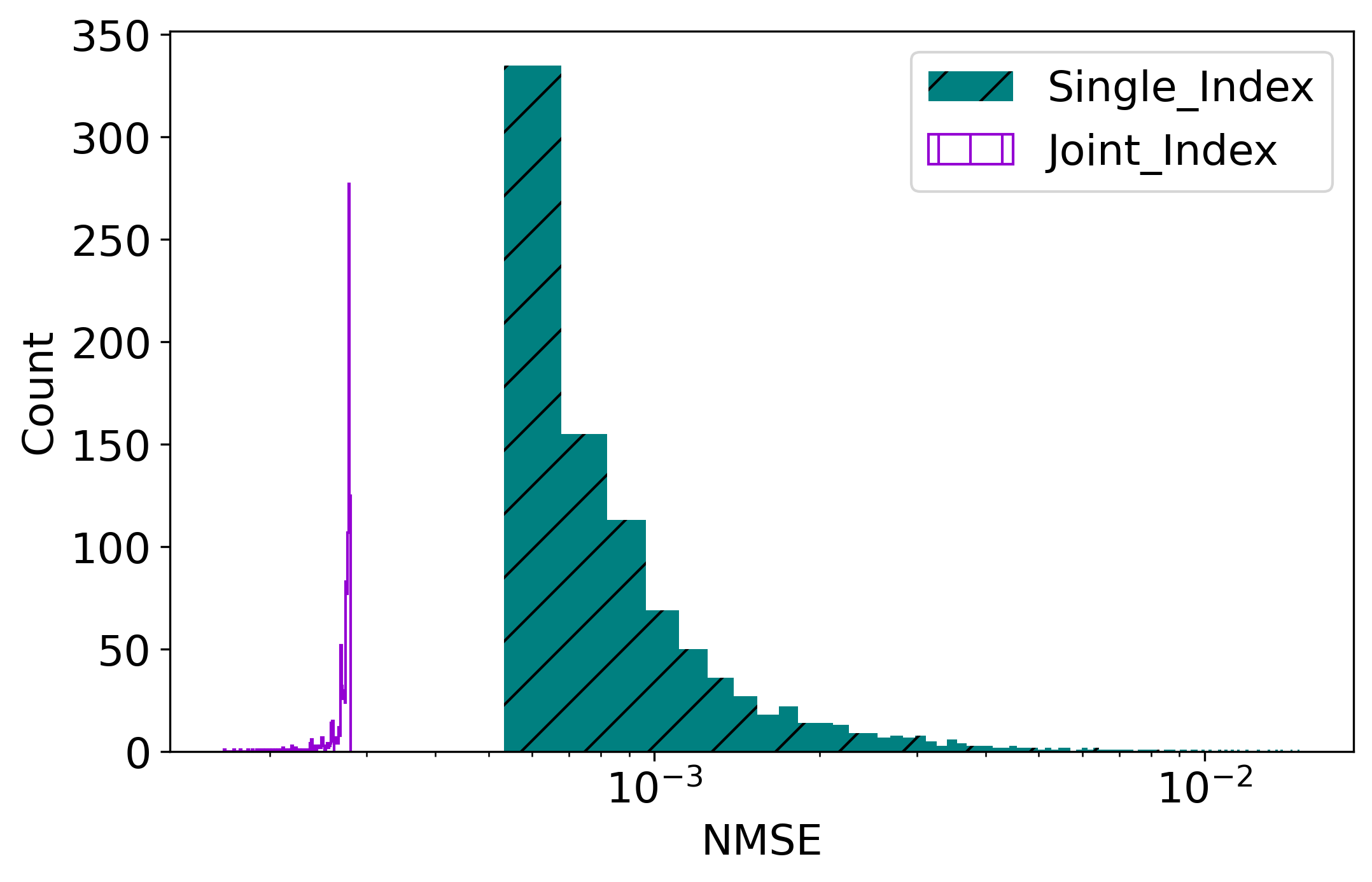}
\caption{NMSE from both single-index (cyan, slant hatched) and joint-index (purple, vertical hatched) estimations considering 1000 noise realizations. These histograms are for spectral indices (0,2/3,3) respectively from left to right. For all these realizations, NMSE for single-index is larger than the joint-index estimation, and NMSE for joint-index is always close to zero. This further verifies the unbiased nature of joint-index multicomponent estimations.}
\label{fig:noise_nmse}
\end{figure*}
%
To prove that the NMSE behavior we showed in Table~\ref{tab:nmse} is not obtaining by chance, we repeat the procedures described in the injection study for 1000 noise realizations. We then compute the NMSE for each case. The histograms obtained from this study employing both single-index and joint-index estimations for all the three spectral indices are shown in Figure~\ref{fig:noise_nmse}. It is evident from the histograms that, for all these cases, NMSE for joint-index estimation is close to zero indicating a better source reconstruction. On the other hand, the single-index estimations are biased, indicated by the scattered nature of NMSE which is always significantly greater than the one from joint-index estimation. This further confirms the better source recovery of the joint-index multicomponent estimations. 

\begin{figure*}
\centering
\includegraphics[width = 0.32\textwidth]{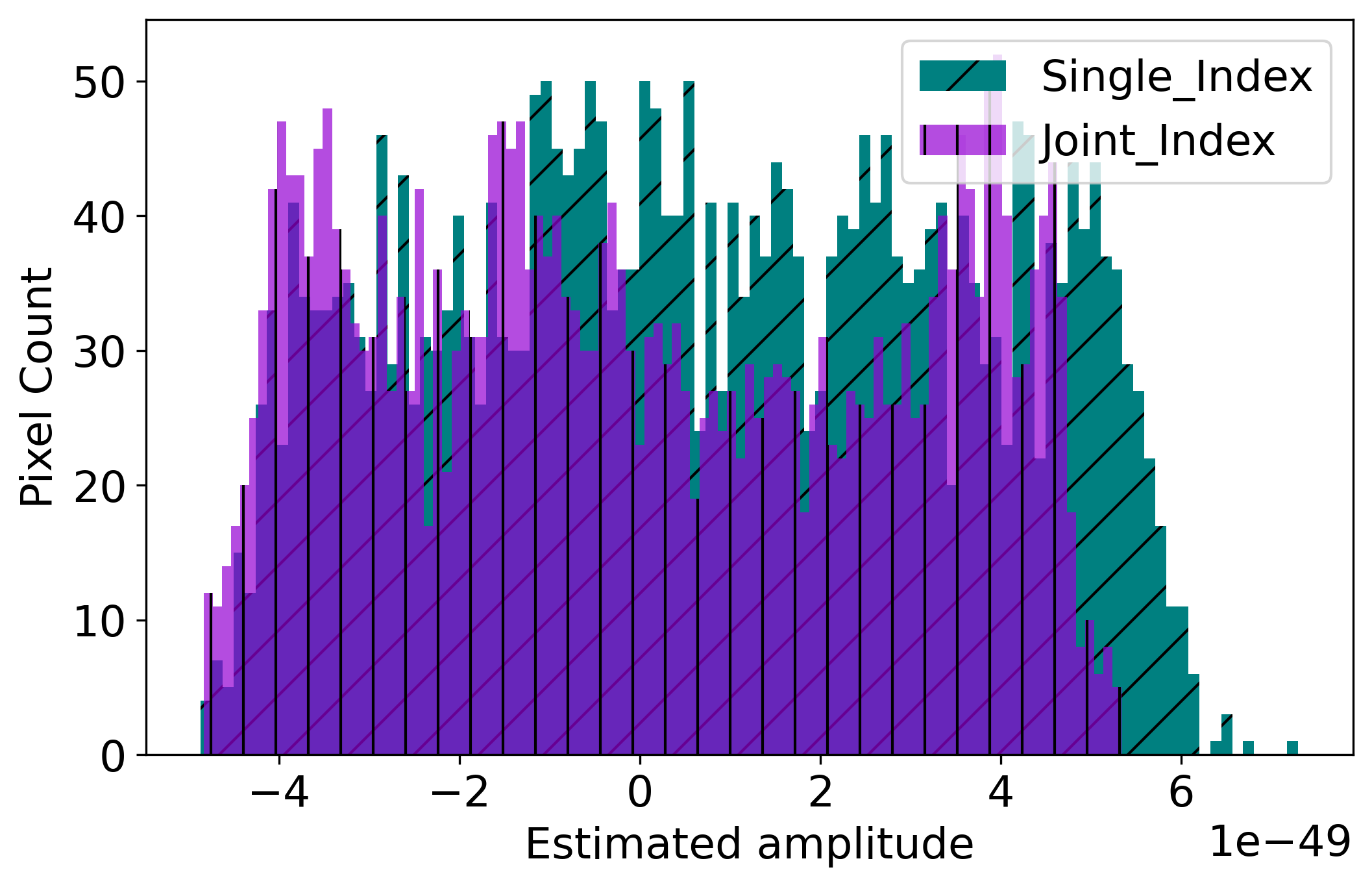}
\includegraphics[width = 0.32\textwidth]{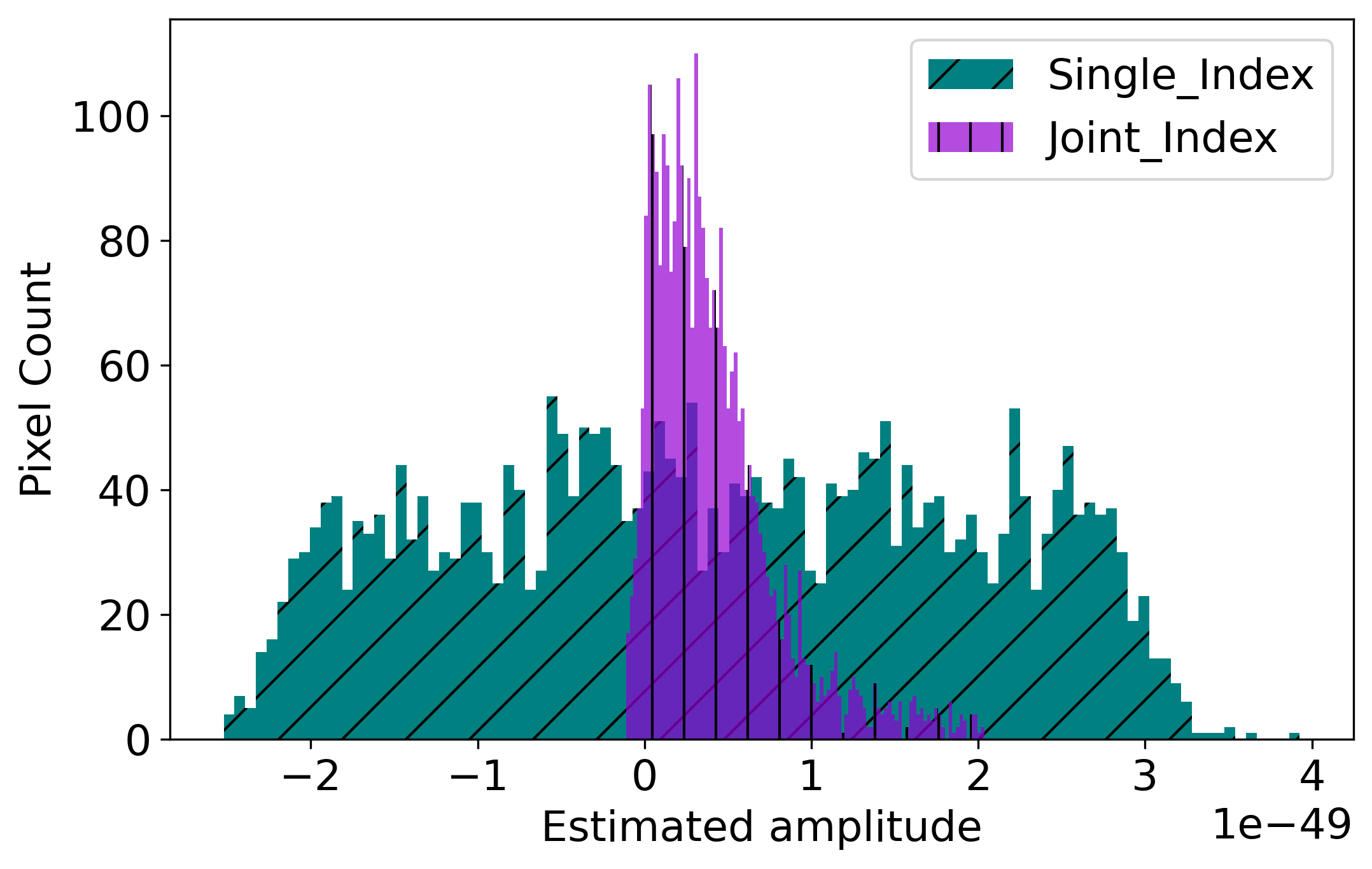}
\includegraphics[width = 0.32\textwidth]{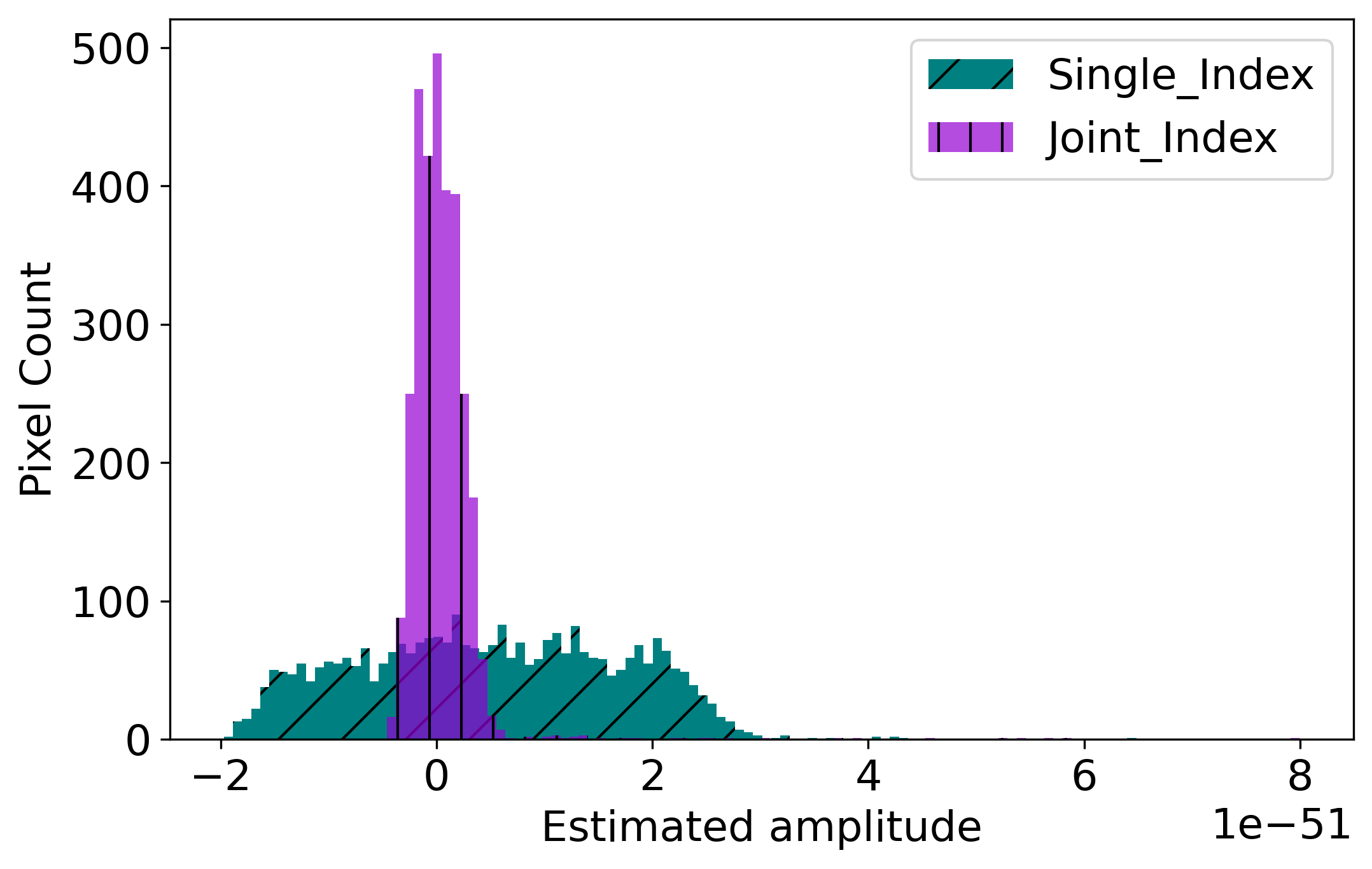}\\
\includegraphics[width = 0.32\textwidth]{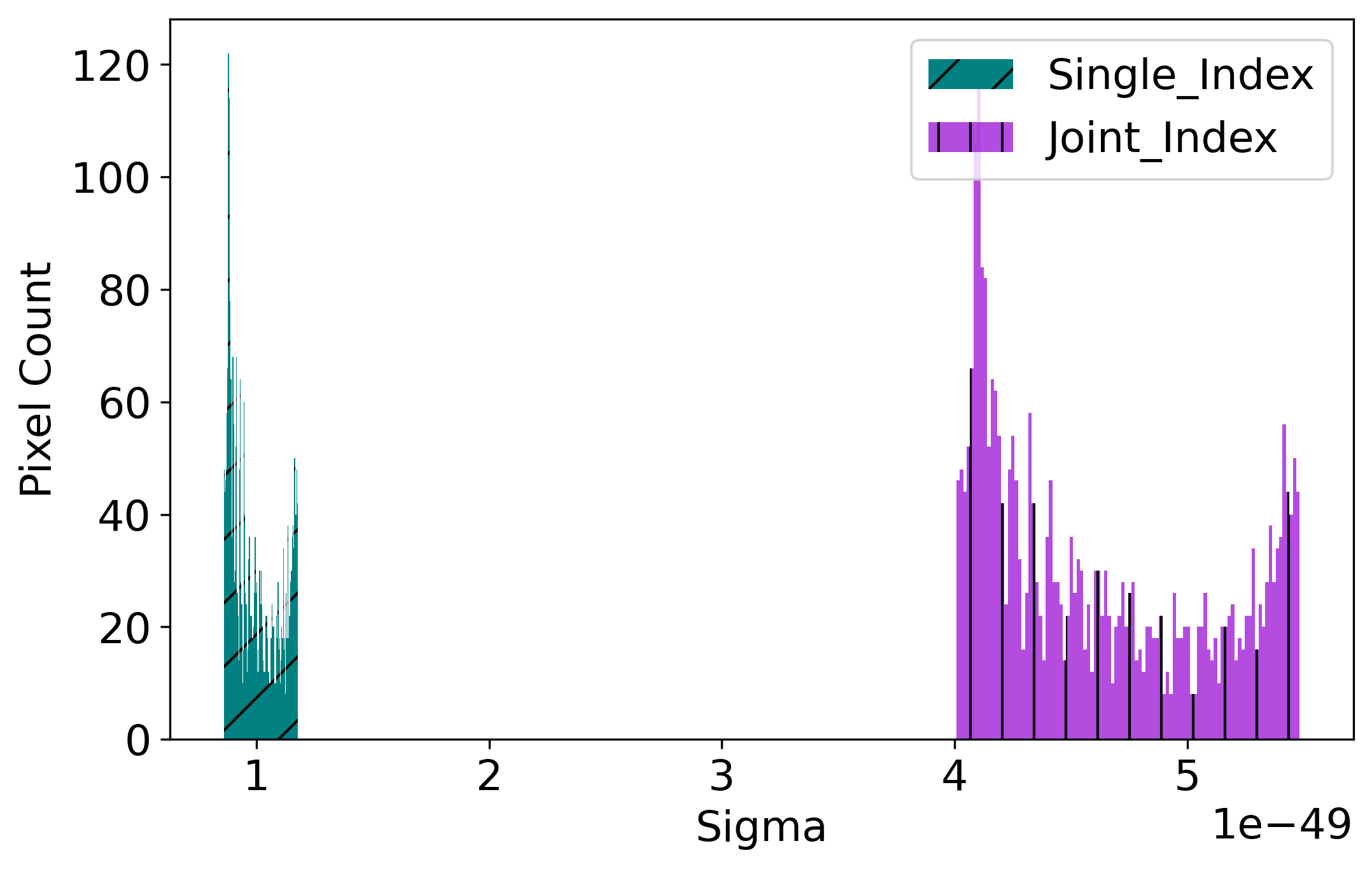}
\includegraphics[width = 0.32\textwidth]{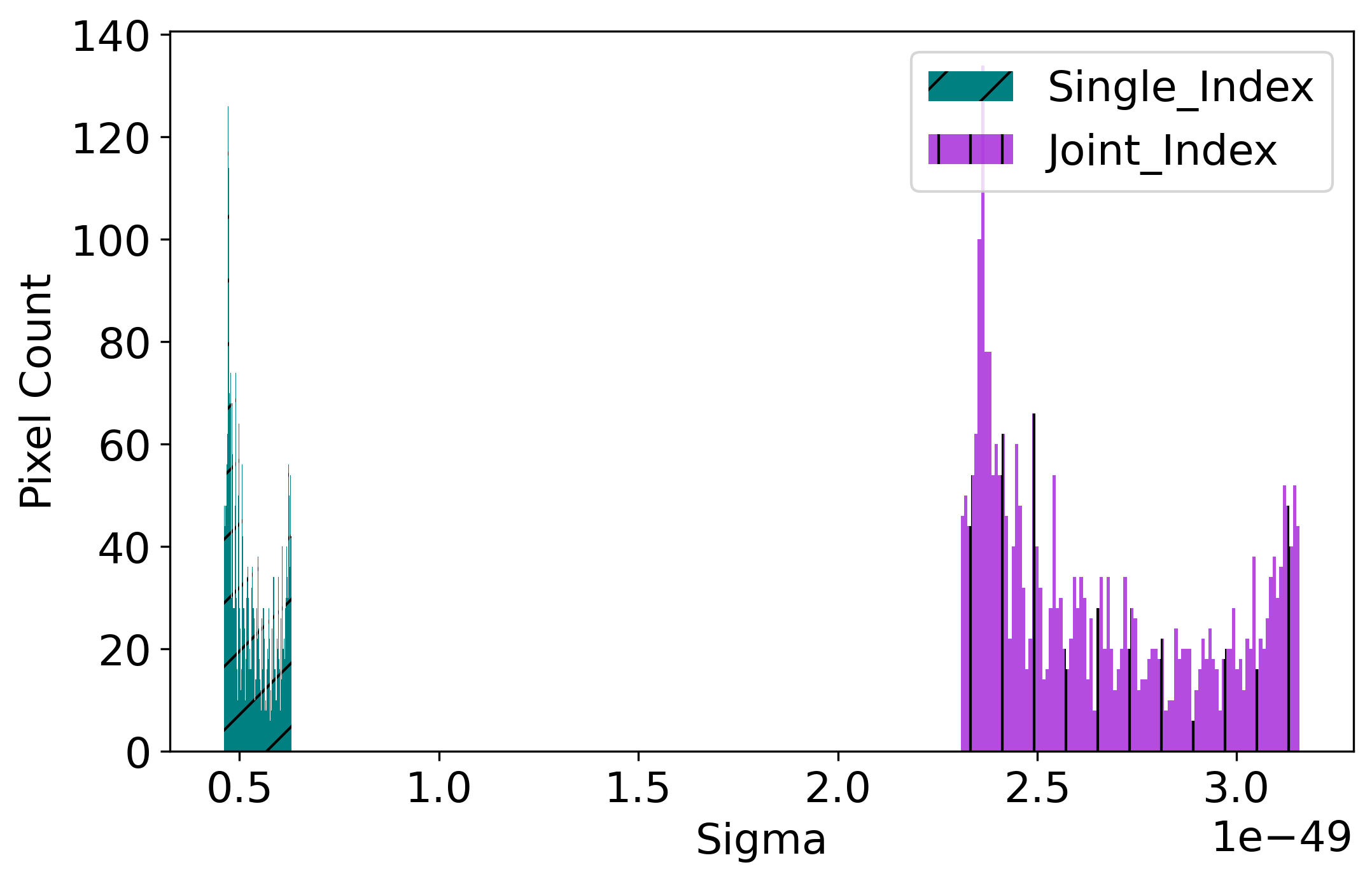}
\includegraphics[width = 0.32\textwidth]{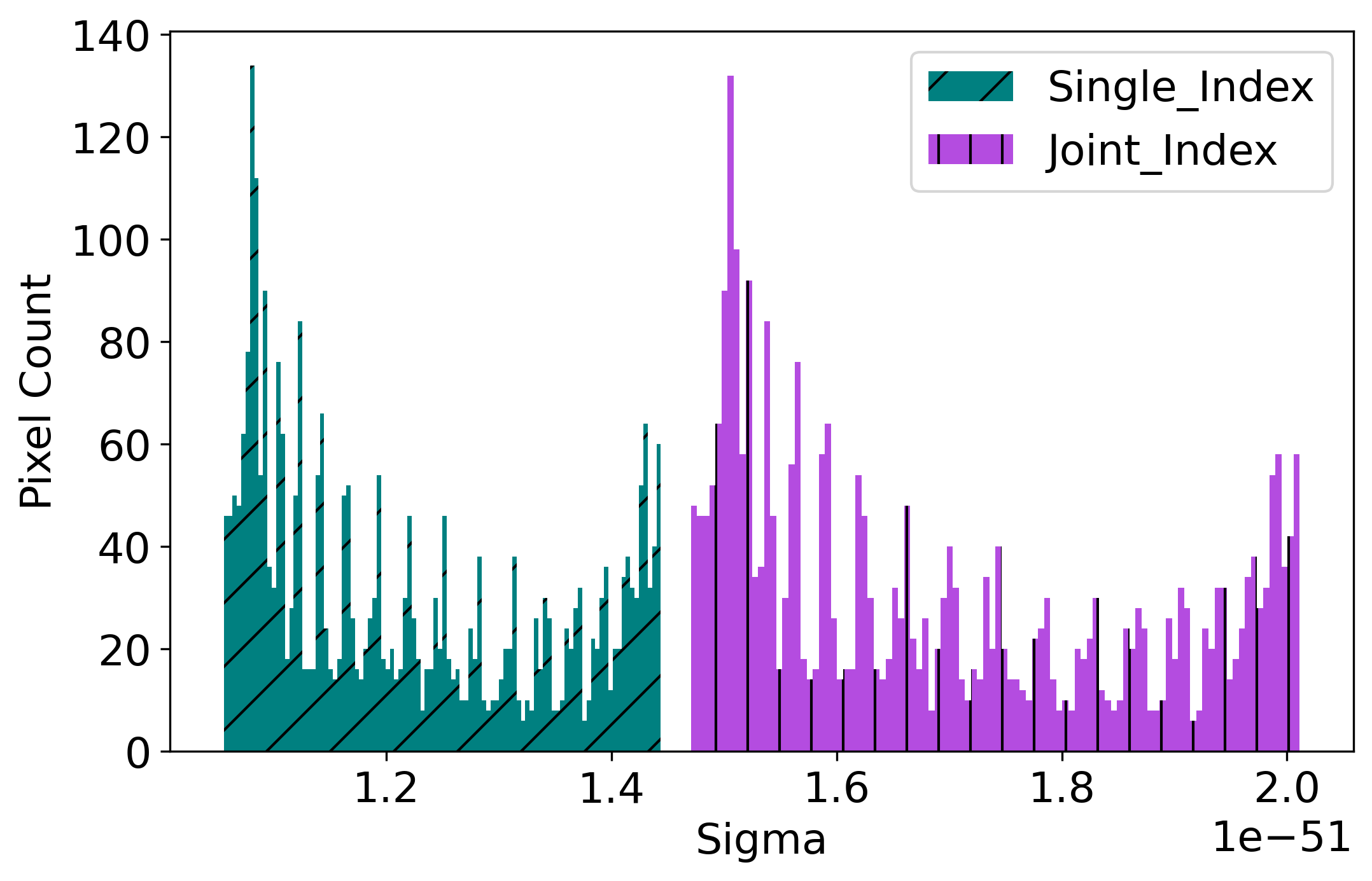}\\
\includegraphics[width = 0.32\textwidth]{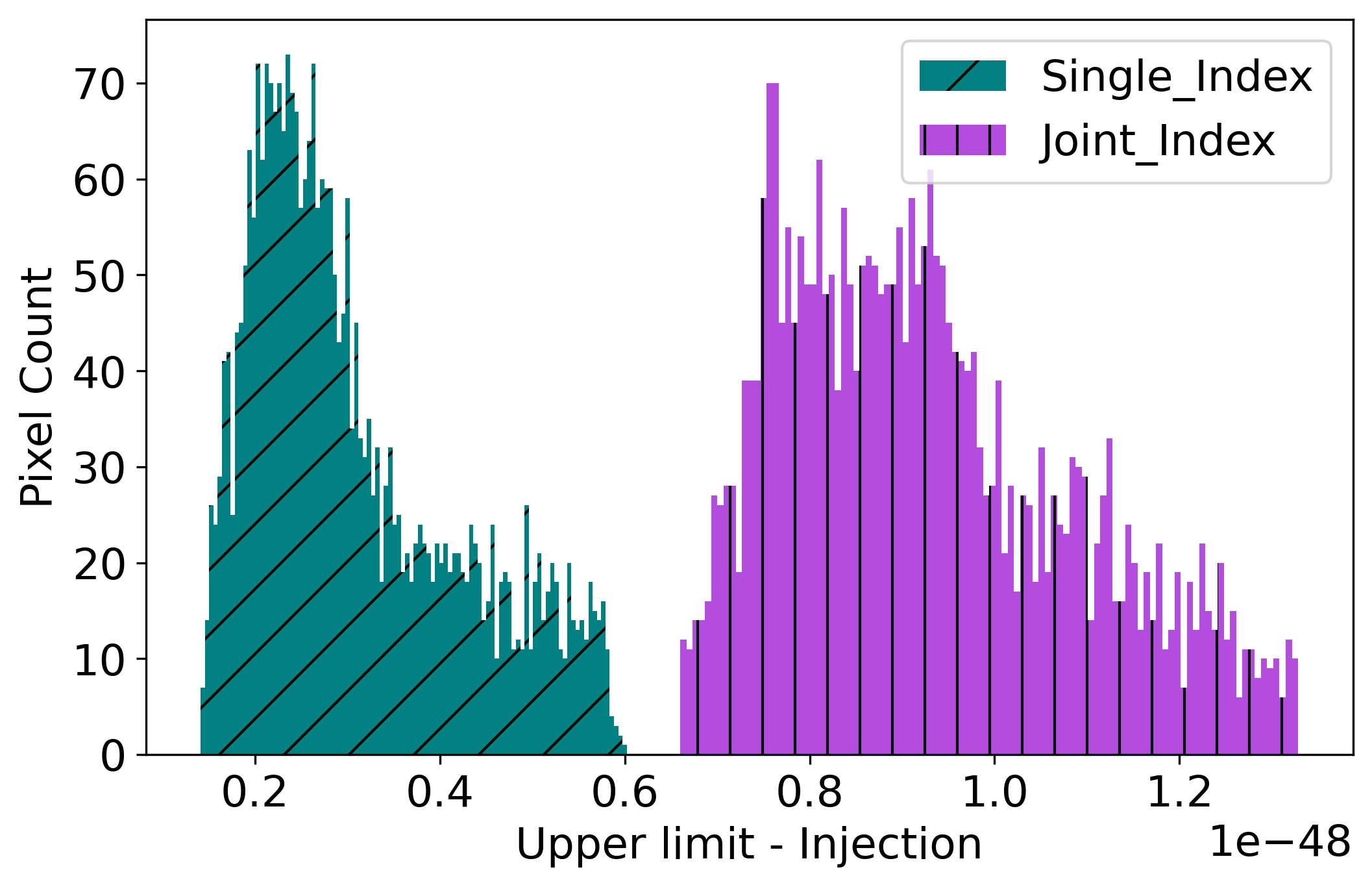}
\includegraphics[width = 0.32\textwidth]{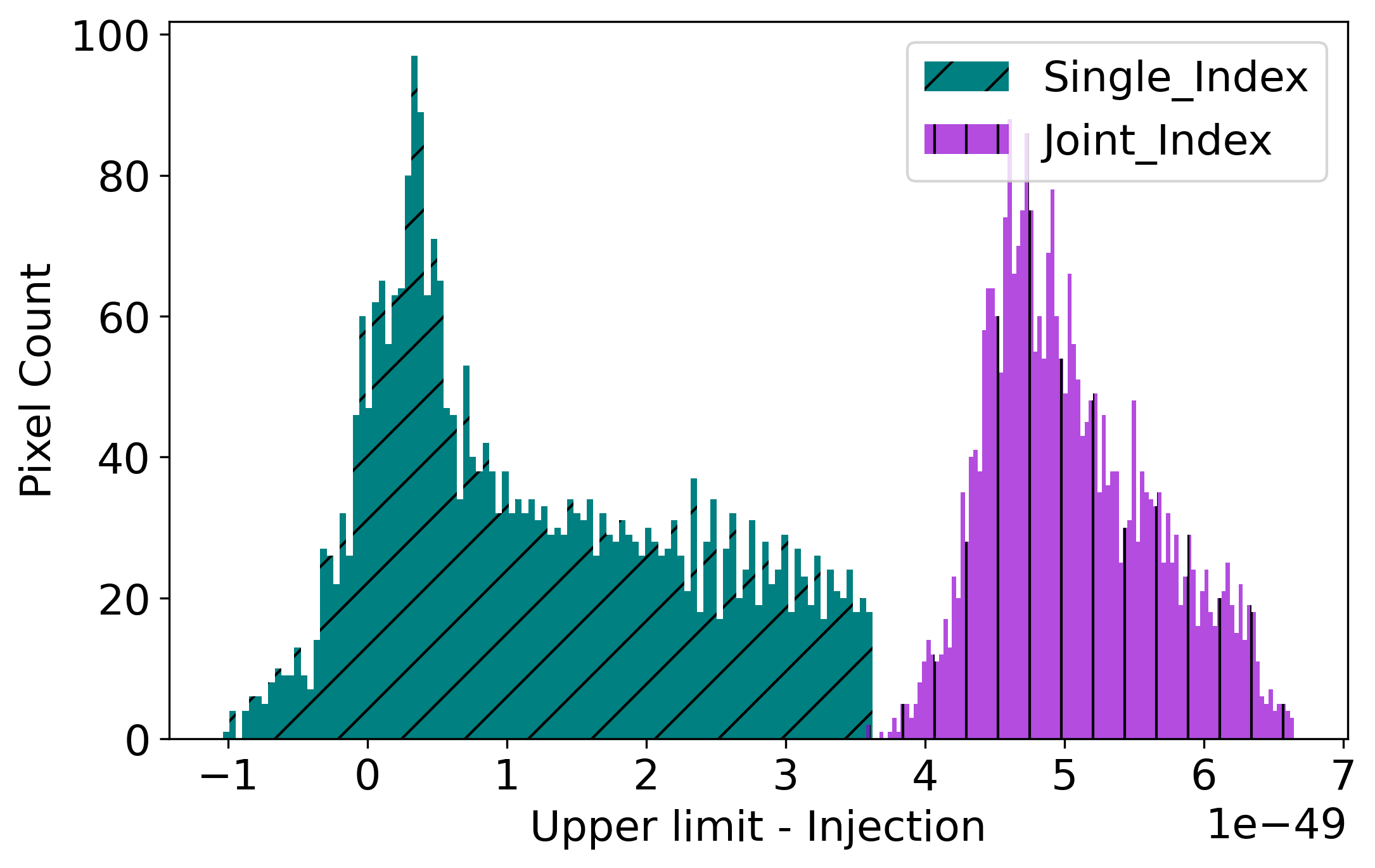}
\includegraphics[width = 0.32\textwidth]{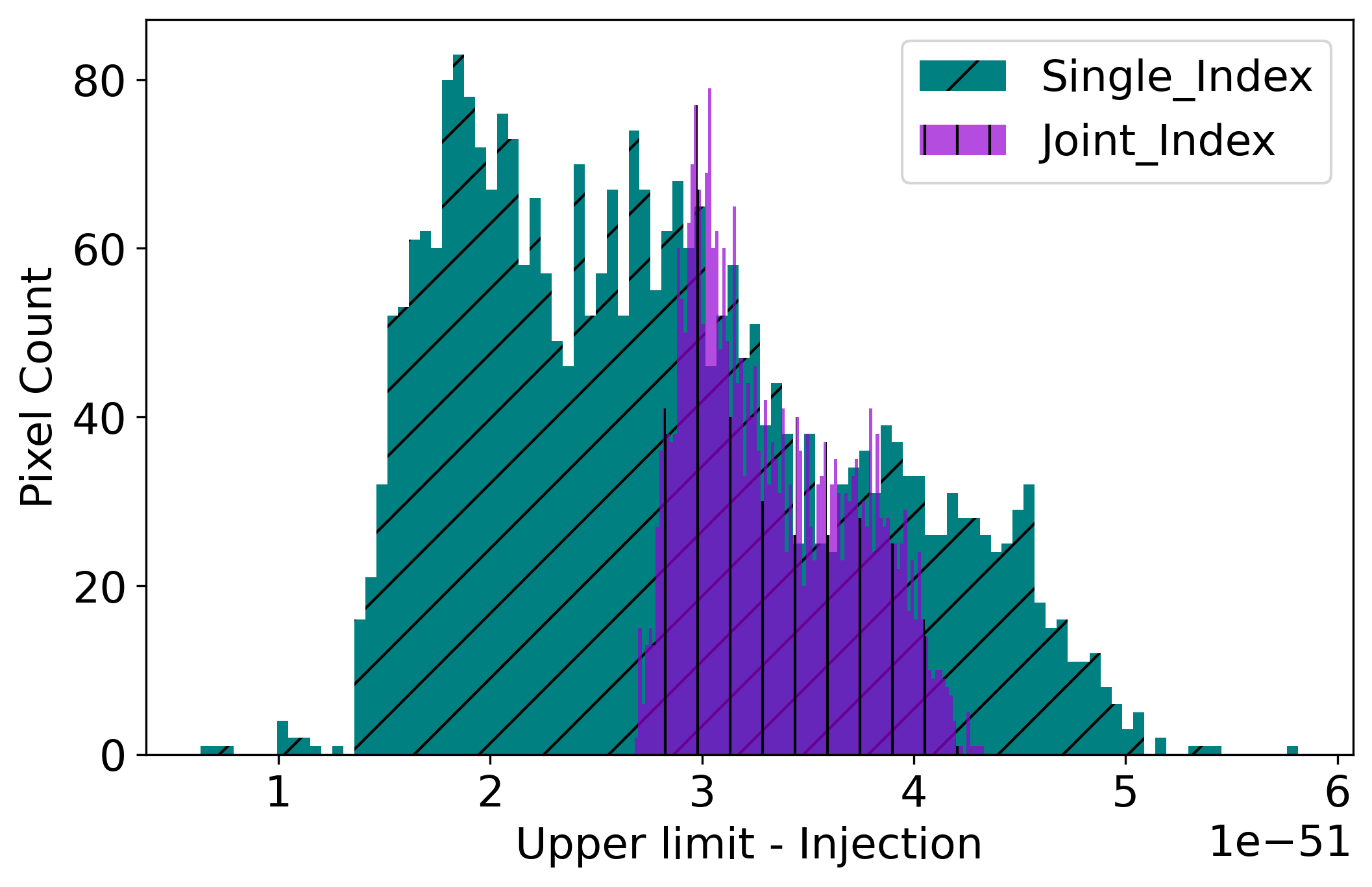}
\caption{Estimated amplitude and $95\%$ upper limit along with the variance of the estimators from the injection study (with corresponding injection amplitudes of $5\times10^{-49}, 2\times10^{-49}$ and  $8\times10^{-51}$ respectively) is shown for $\alpha = 0,2/3,3$ from left to right respectively. In all the histograms, cyan (slant hatched) represents the single-index analysis and purple (vertical hatch) shows the joint-index analysis. Top panel histograms show the estimated amplitudes from both the methods. Middle panel depicts the variances of single-index and joint-index estimators. The bottom panel shows the differences between the estimated upper limits and the (high SNR) injections. For a valid 95\% upper limit, the difference should be positive for more than $\sim 95$\% pixels. However, this is not the case for the single index analysis for $\alpha = 2/3$, as seen in the middle column of the bottom panel, where $\sim 11\%$ of the pixels have negative differences.}
\label{fig:ul_hist}
\end{figure*}

\begin{figure*}
\centering
\includegraphics[width = \columnwidth]{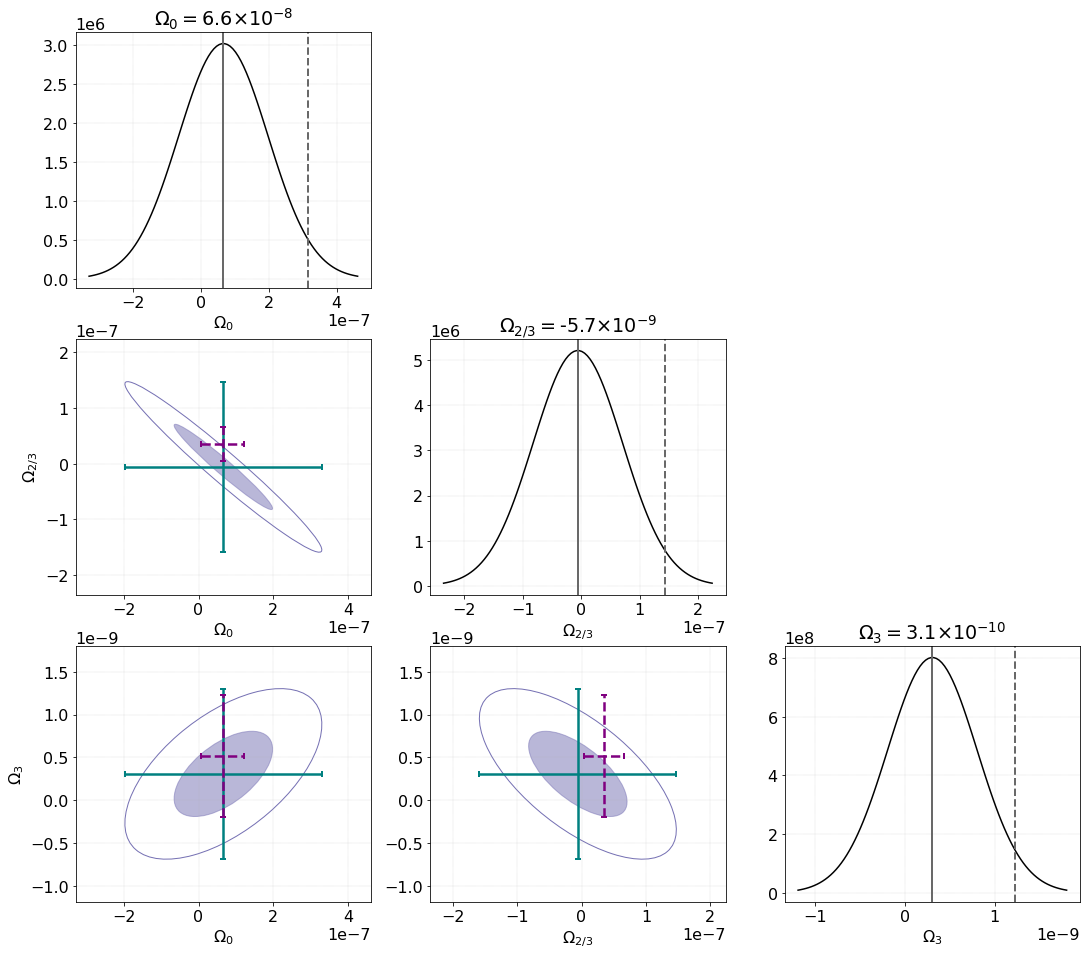}
\includegraphics[width = \columnwidth]{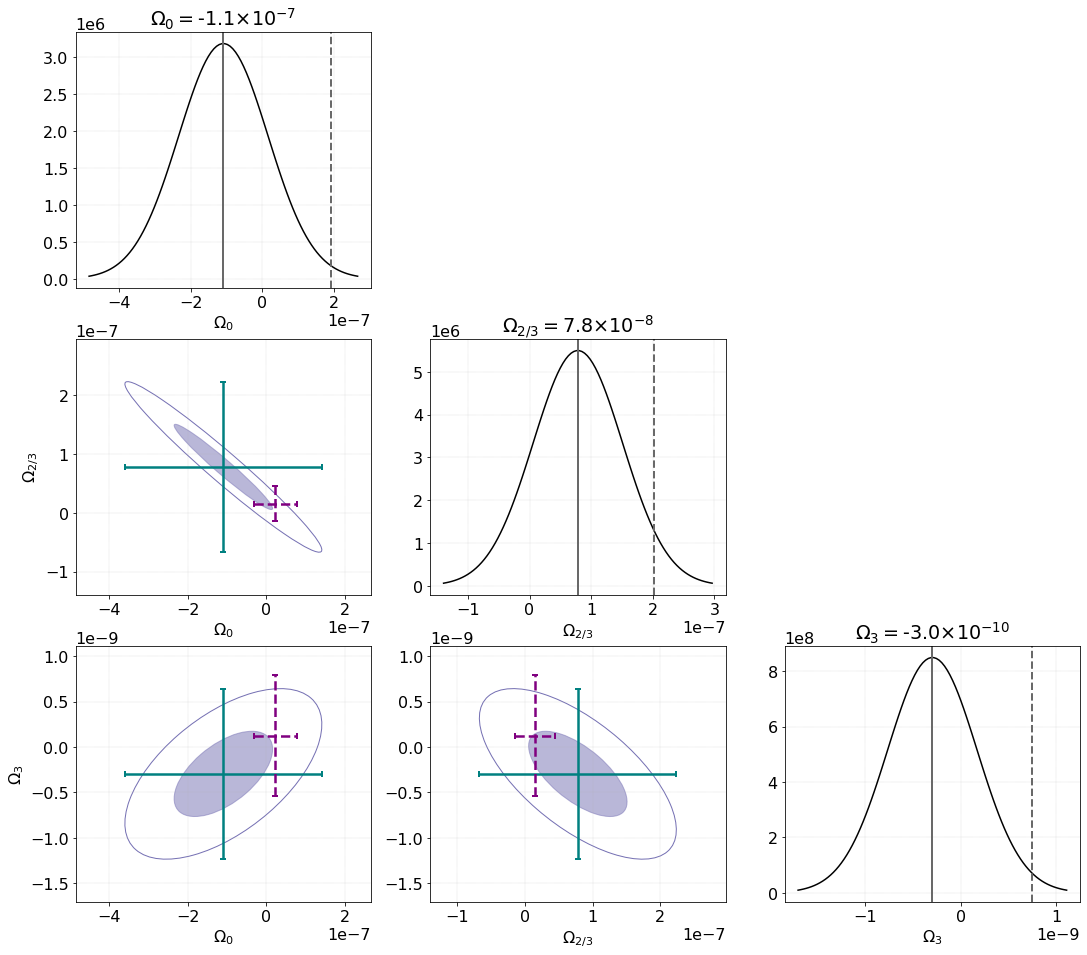}\\
\includegraphics[width = \columnwidth]{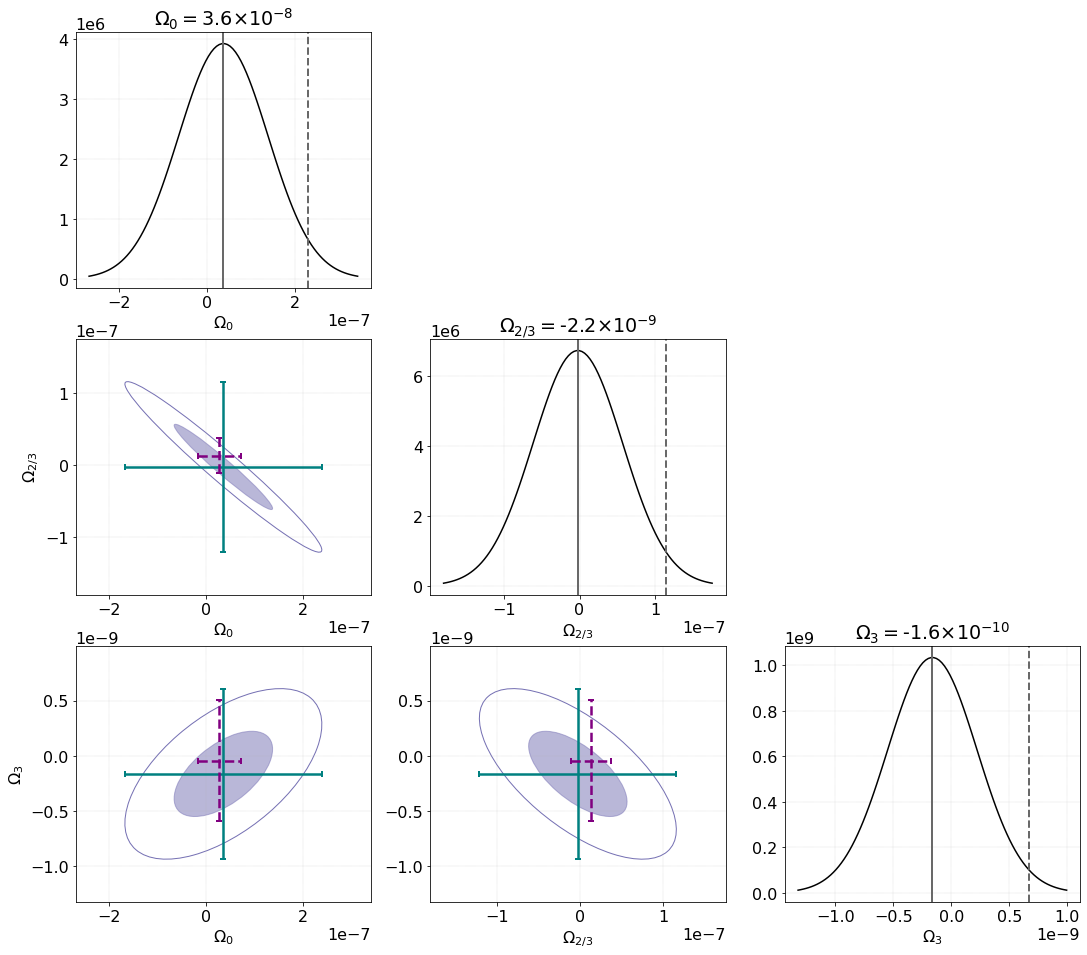}
\includegraphics[width = \columnwidth]{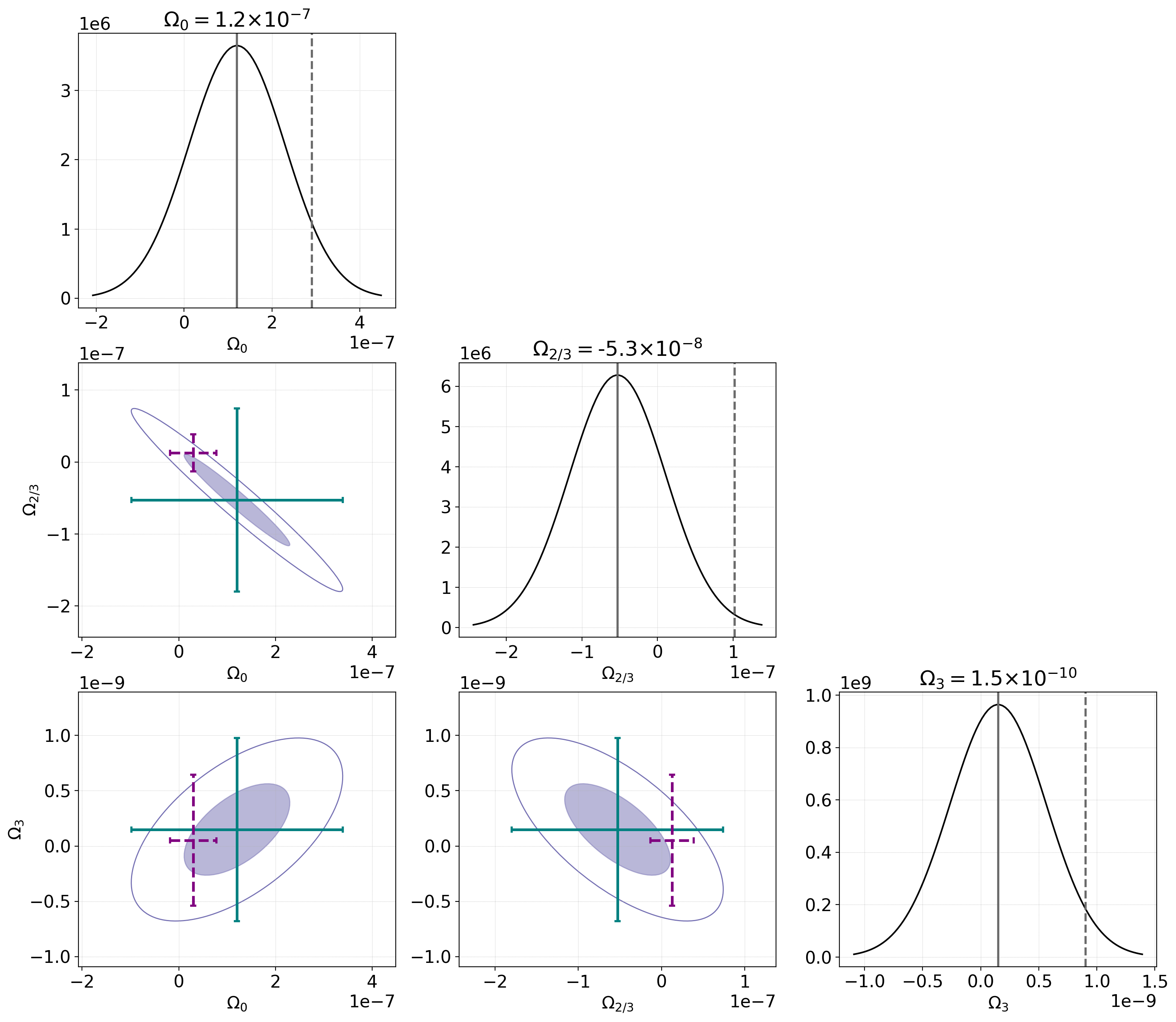}
\caption{The estimates of three SGWB components (for $\alpha = 0,2/3,3$) using O1+O2+O3a folded dataset for four random sky directions (pixel number =  [71,200,1566,1999] in a \hpx resolution of $N_{\mbox{side}} = 16$ following the RING~\cite{HEALPix} ordering scheme) are shown. The purple (dashed) $2\sigma$ error bar shows the result from a single-index analysis, whereas the cyan (solid) one represents the error bar from the joint-index analysis. The shaded and unshaded regions represent the $1\sigma$ and $2\sigma$ confidence intervals respectively. The one-dimensional plot depicts the marginalized probability distribution for each component. Here the solid vertical line denotes the mean, and the dotted vertical line represents the $95\%$ confidence upper limit after marginalizing over the calibration uncertainty~\cite{Whelan_2014} and assuming a uniform prior on SGWB flux. As evident from Figure~\ref{fig:ul_hist}, when there is an injection, the intersection of cyan (solid) error bars pass through the point marking the injection value. The plot establishes that the effect of covariance between the spectral shape on the estimates are significant and hence the need for a multicomponent analysis. }
\label{fig:ul_corner}
\end{figure*}

Given the current sensitivity levels of the detectors, when we analyze real data, we will be more interested in setting an upper limit on the SGWB components. Here, we investigate the effect of the joint-index multicomponent separation method on the upper limit. We first consider three spectral shapes and three sky patterns as in Fig.~\ref{fig:injection} and use the same injection strength. We then compute the $95\%$ confidence upper limit~\cite{Whelan_2014,S5_Dir} on the amplitude of the recovered signals from both methods. Fig.~\ref{fig:ul_hist} show the results obtained from this study. In the top panel, the estimated amplitudes from both single-index and joint-index estimations are shown. Comparing with the injection strength, single-index estimations always overestimates the amplitude of the individual components. We explore further to see the implication of a biased amplitude estimation on the upper limit. We first compute the variance of single and joint-index estimators. As we noted earlier, the variance or the error bar is underestimated for the single-index analysis. From earlier studies~\cite{Parida_isotropic,O1O2Folded,Eric09}, it is shown that when the parameters of the analysis increase, the variance of the corresponding estimator also increases (see Appendix~\ref{sec:appendix} for more details). Here, for the joint-index analysis, we observe the same behavior. These responses are depicted with the aid of histograms in the middle panel of Fig.~\ref{fig:ul_hist}. We then compute the $95\%$ confidence upper limit at every pixel on the sky using both methods. The results are shown in the bottom panel of Fig.~\ref{fig:ul_hist}. The histogram in the bottom panel shows the difference between the upper limit we computed and the injection strength. However, if we consider, $\alpha = 2/3 $ case in our injection study, we can see that a considerable amount ($\sim 11\%$) of the points in the histogram fall under the negative region. This indicates that, even though the single-index estimate appears to provide a stronger upper limit, it violates the requirement for a 95\% confidence upper limit, at least in certain generic cases, where for more than 5\% pixels ($\sim 11$\% here) the upper limit is lower than the injected value (To be more rigorous, one should create several signals and noise realizations and find out in what fraction of cases the upper limits are lower than the injected value. This will however be computationally expensive. If one assumes that most pixels are reasonably independent, one would expect $95\%$ of the pixels to have upper limits higher than the injected value). This exercise suggests that even if the detectors are not sensitive enough to detect SGWB, the joint-index multicomponent estimator provides safer upper limits when one cannot ignore the existence of more than one component.

\section{Results and Conclusion}
\label{sec:result}
\begin{figure*}
\centering
\includegraphics[width = 0.32\textwidth]{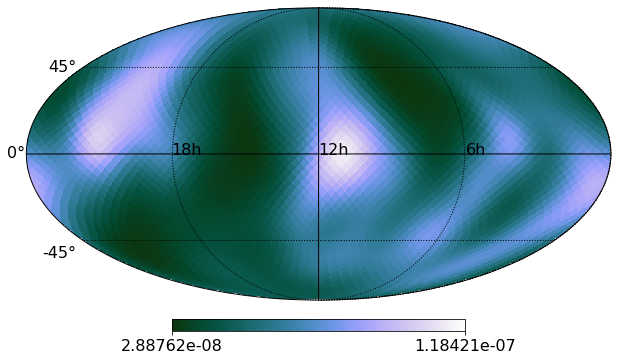}
\includegraphics[width = 0.32\textwidth]{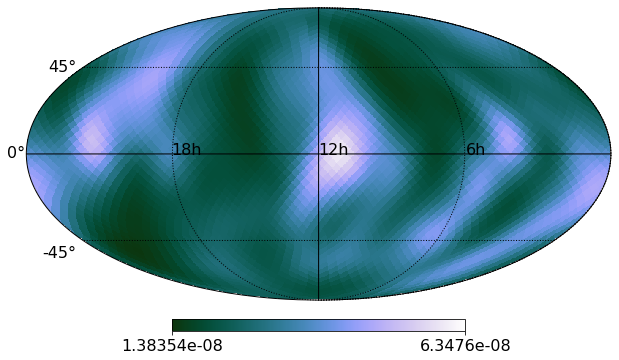} 
\includegraphics[width = 0.32\textwidth]{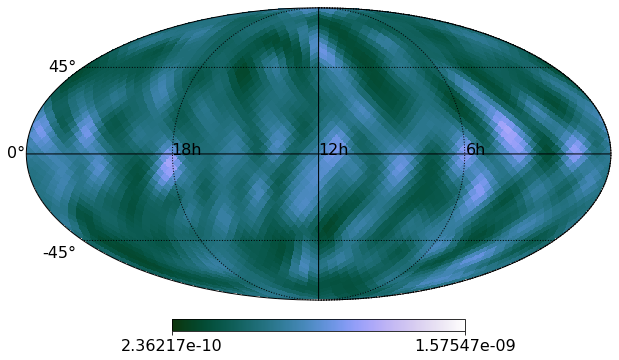} \\
\includegraphics[width = 0.32\textwidth]{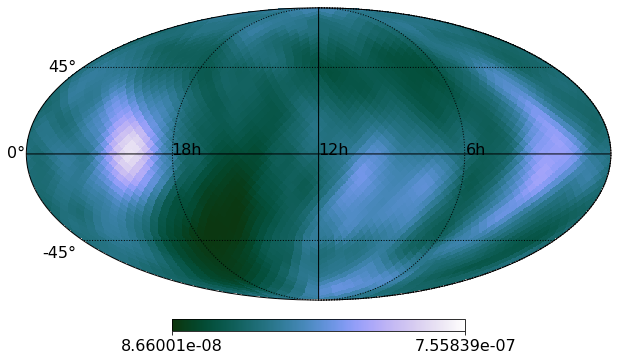}
\includegraphics[width = 0.32\textwidth]{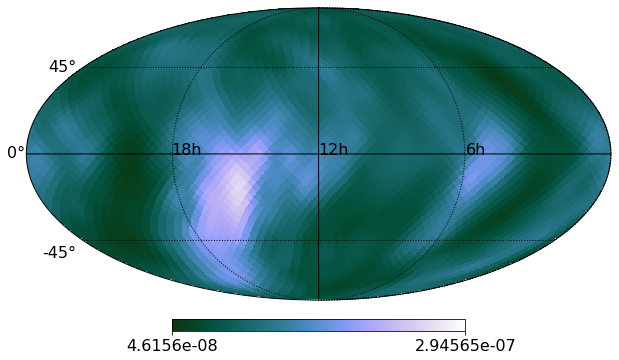} 
\includegraphics[width = 0.32\textwidth]{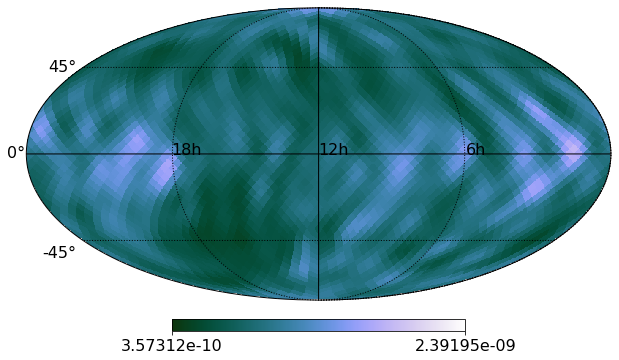}
\caption{Sky maps representing the $95\%$ upper limit on the SGWB energy flux (${\rm erg \, cm^{-2} \, s^{-1} \, Hz^{-1}}$) obtained from O1+O2+O3a folded dataset considering three spectral index jointly.  Top panel shows the upper limit from the single-index component separation whereas the bottom panel denotes the joint-index multicomponent estimation. From left to right the spectral index corresponding to the sky maps are $\alpha=0,2/3,3$. All the maps are represented as a color bar plot on a Mollweide projection of the sky in ecliptic coordinates. For the visualization purpose we smooth the upper limit sky maps with a 10 degree Gaussian beam and adjusted the color bar scale using the values from Table~\ref{tab:upperlimit}}
\label{fig:upper_limit}
\end{figure*}
\begin{figure*}
\centering
\includegraphics[width = 0.32\textwidth]{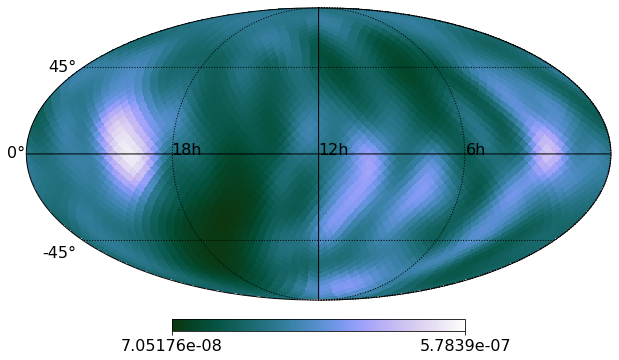}
\includegraphics[width = 0.32\textwidth]{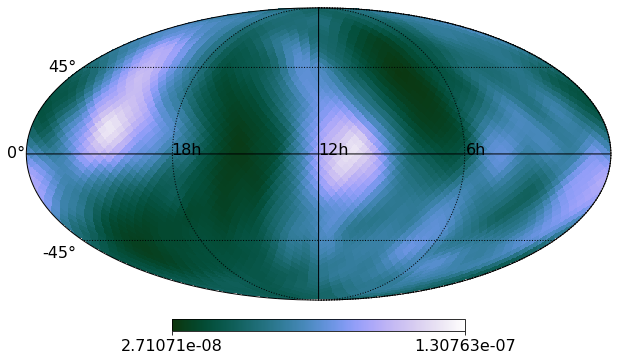}
\includegraphics[width = 0.32\textwidth]{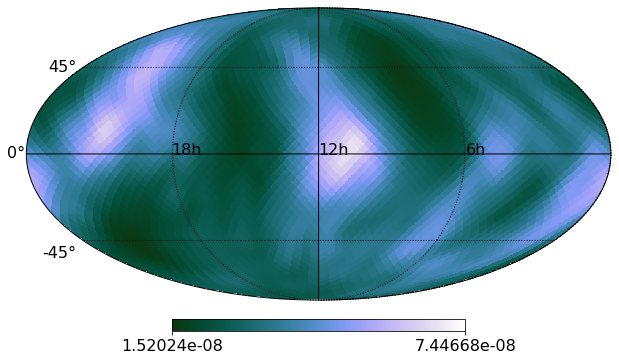}  \\
\includegraphics[width = 0.32\textwidth]{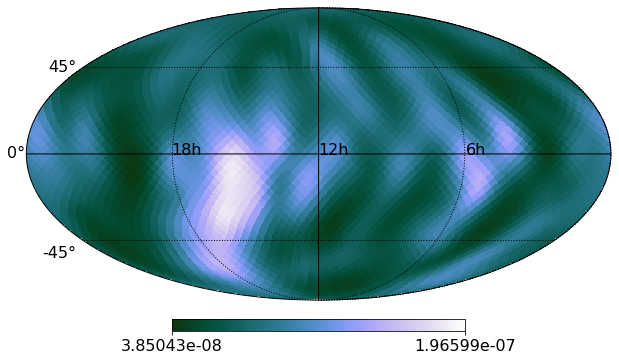} 
\includegraphics[width = 0.32\textwidth]{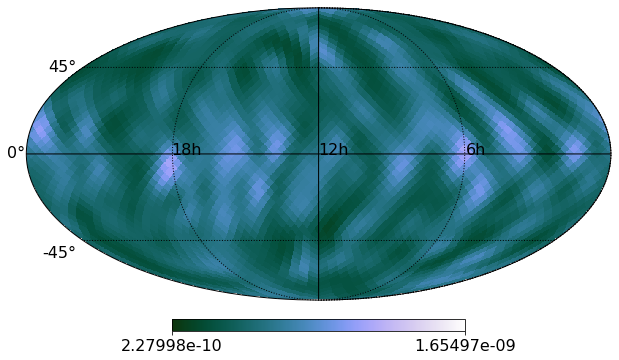}
\includegraphics[width = 0.32\textwidth]{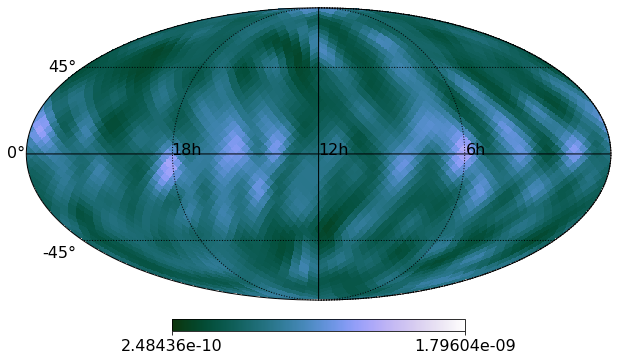}
\caption{Sky map representing $95\%$ upper limit on the SGWB energy flux (${\rm erg \, cm^{-2} \, s^{-1} \, Hz^{-1}}$) obtained from O1+O2+O3a folded dataset by jointly considering two spectral index. From left to right each column represent upper limit sky maps from joint-index analysis considering (0,2/3), (0,3), and (2/3,3) spectral index combinations. All the maps are represented as a color bar plot on a Mollweide projection of the sky in ecliptic coordinates. For the visualization purpose we smooth the upper limit sky maps with a 10 degree Gaussian beam and adjusted the color bar scale using the values from Table~\ref{tab:upperlimit}.}
\label{fig:upper_limit_2index}
\end{figure*}

We now apply the joint-index multicomponent separation method to the folded dataset derived using the observational data from aLIGO’s H and L detectors, taken up to the first half of the third observing run. Similar to what we did for the injection study, we first compute the dirty maps from the O1+O2+O3a observation data by employing the optimal filtering~\cite{allen-romano}. We then construct the coupling matrix by assuming three spectral indices $\alpha = 0, 2/3, 3$ from the same dataset. To obtain the true SGWB sky, we solve the matrix equation given in Eq.~(\ref{eq:matrix_convolution}). Here we account for spectral index covariance. 

Since the noise level from the observed data is still larger for signal detection, the $95\%$ upper limits are placed on the gravitational wave energy flux (units of ${\rm erg \, cm^{-2} \, s^{-1} \, Hz^{-1}}$)~\cite{S5_Dir,O1paper,O2paper,o3_direction}. To further demonstrate the importance of a joint-index estimation, we choose four pixels corresponding to four random sky directions and explore the effect of the covariance between the spectral indices.
We compute both single and multicomponent upper limits for these directions (given the dataset is dominated by noise and the existence of pixel-to-pixel covariance, it is not easy to show the corner plot for every pixel). The one-dimensional marginalized probability distribution and the error ellipses are shown in the corner plot Fig.~\ref{fig:ul_corner}. These corner plots show that the upper limit gets underestimated for different pixels in the sky if we rely on single-index analysis. We can also see that this deviation, compared to the joint-index upper limit, is substantial in some cases (see top or bottom right corner plot). From this figure, it is evident that the effect of spectral covariance on the estimation of SGWB components is significant. Given the results presented here, in order to account for these effects, it demands the use of joint-index multicomponent estimation in the SGWB search. 

In Fig.~\ref{fig:upper_limit}, we provide the $95\%$ upper limit range for the gravitational wave energy flux for all the sky directions. The top panel shows the upper limit sky map obtained using the single-index estimation. Whereas the joint-index multicomponent estimation is shown in the bottom panel. All these sky maps are consistent~\footnote{It is not straightforward to exactly compare the results shown here with the LVK results in \cite{o3_direction} as the datasets are slightly different.} with the upper limit range quoted in the most recent directional limits from the LVK collaboration~\cite{o3_direction}. We also showed the $95\%$ credible upper limit sky maps by simultaneously considering two spectral indices in Figure~\ref{fig:upper_limit_2index}. Here each column represents the results from all three combinations of spectral indices. We summarize these results in Table~\ref{tab:upperlimit} for easy comparison. Here the upper limits on individual source components are placed following the single-index estimation, the SGWB amplitudes are overestimated and hence biased. Therefore, the unbiased results from the joint-index estimation give a more accurate representation of the SGWB background upper limit from individual source categories if non-negligible multiple components are present together. 

\begin{table}
    \centering
 \begin{tabular}{c|c|c|c|c|c}
 \hline\hline
\multicolumn{6}{c}{95\% Upper Limit $\times 10^{-8}$} \\
\hline
\multirow{3}{*}{$\alpha$} & \multirow{3}{*}{Single Index} & \multicolumn{4}{c}{Joint Index}                                    \\ \cline{3-6} 
                          &                               & \multicolumn{3}{c|}{two-$\alpha$} & \multirow{2}{*}{three-$\alpha$} \\ \cline{3-5}
                          &                               & 0 , 2/3     & 0 , 3       & 2/3 , 3     &                                 \\ \hline
0                         & 2.89-11.8                    & 7.05-57.84 & 2.71-13.07 &           & 8.66-75.58                      \\ \hline
2/3                       & 1.38-6.35                     & 3.85-19.66 &           & 1.52-7.45 & 4.62-29.45                      \\ \hline
3                         & 0.024-0.157                     &           & 0.023-0.165 & 0.025-0.18 & 0.036-0.24                      \\ \hline \hline
\end{tabular}
    \caption{$95\%$ Upper limit range across all the sky direction, on the gravitational wave energy flux (${\rm erg \, cm^{-2} \, s^{-1} \, Hz^{-1}}$) from single-component and joint-index multicomponent estimations (simultaneously considering two and three spectral shapes) using the O1+O2+O3a folded dataset. Upper limit sky maps for each cases are shown in Figure~\ref{fig:upper_limit} and \ref{fig:upper_limit_2index}. The increase in upper limit exhibited by the joint-index estimation is due to the increase in the variance of the corresponding estimator.}
     \label{tab:upperlimit}
\end{table}

Relating the SGWB signal to the sources that contribute to it is a challenging problem. While extracting the individual source contribution, minimizing the bias in the estimators is important. With the increase in detector sensitivity, an unbiased estimator, like the one shown in this paper will act as an essential tool. The method and results shown in this work will set a benchmark for this future problem, which is likely to become an essential task after the first detection of SGWB.

This analysis can be extended to multiple directions. Since the now-standard analysis pipeline {\tt{PyStoch}} can perform a model-independent multi-baseline search for the anisotropic SGWB straightforwardly, it will be interesting to combine it with the proposed method. This will enable us to simultaneously estimate the spectral, frequency, and angular dependence of any SGWB signal. This will be the keystone to future component separation efforts. It will be also interesting to look at this component separation problem using the spherical harmonic (SpH) decomposition techniques. Given the information about the coupling of different SpH modes are encoded in the SpH fisher information matrix, it could potentially help us in separating the different spectral shape contributions to the SGWB backgrounds from different multipoles. At this stage, since the LVK stochastic analysis set the upper limit assuming three source categories, characterized by three spectral shapes, the component separation method presented in this paper will augment the existing analysis.

\begin{acknowledgments}
The authors thank Patrick Meyers for carefully reading the manuscript and providing valuable comments. This work significantly benefited from the interactions with the Stochastic Working Group of the LIGO-Virgo-KAGRA Scientific Collaboration. We acknowledge the use of IUCAA and Caltech LDAS clusters for the computational/numerical work. J. S. acknowledges the support by Japan Society for the Promotion of Science (JSPS) KAKENHI Grant No. JP17H06361. S. M. acknowledges support from the Department of Science and Technology (DST), India, provided under the Swarna Jayanti Fellowships scheme. This material is based upon work supported by NSF’s LIGO Laboratory which is a major facility fully funded by the National Science Foundation. This research has also made use of data obtained from the Gravitational Wave Open Science Center (https://www.gw-openscience.org/ ), a service of LIGO Laboratory, the LIGO Scientific Collaboration, and the Virgo Collaboration. LIGO Laboratory and Advanced LIGO are funded by the United States National Science Foundation (NSF) as well as the Science and Technology Facilities Council (STFC) of the United Kingdom, the Max-Planck-Society (MPS), and the State of Niedersachsen/Germany for support of the construction of Advanced LIGO and construction and operation of the GEO600 detector. Additional support for Advanced LIGO was provided by the Australian Research Council. Virgo is funded, through the European Gravitational Observatory (EGO), by the French Centre National de Recherche Scientifique (CNRS), the Italian Istituto Nazionale di Fisica Nucleare (INFN), and the Dutch Nikhef, with contributions by institutions from Belgium, Germany, Greece, Hungary, Ireland, Japan, Monaco, Poland, Portugal, Spain. This paper is assigned the LIGO document control number LIGO-P2100169. Parts of the results in this work make use of the colormaps from the CMasher package~\cite{cmasher_cmap}
\end{acknowledgments}

\appendix
\section{Number of components and variance}
\label{sec:appendix}
\begin{figure*}
    \centering
    \includegraphics[width = 0.32\textwidth]{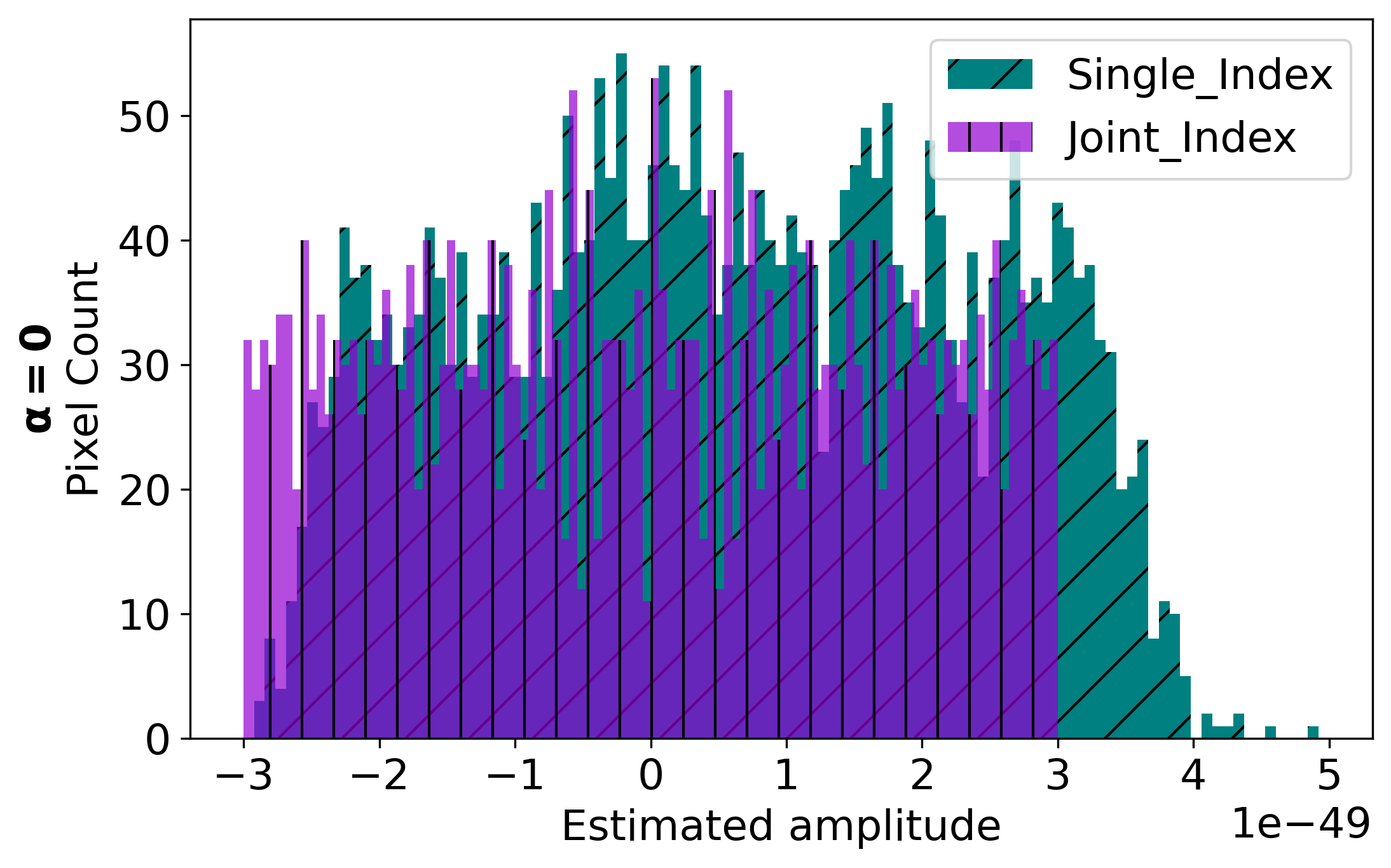}
    \includegraphics[width = 0.32\textwidth]{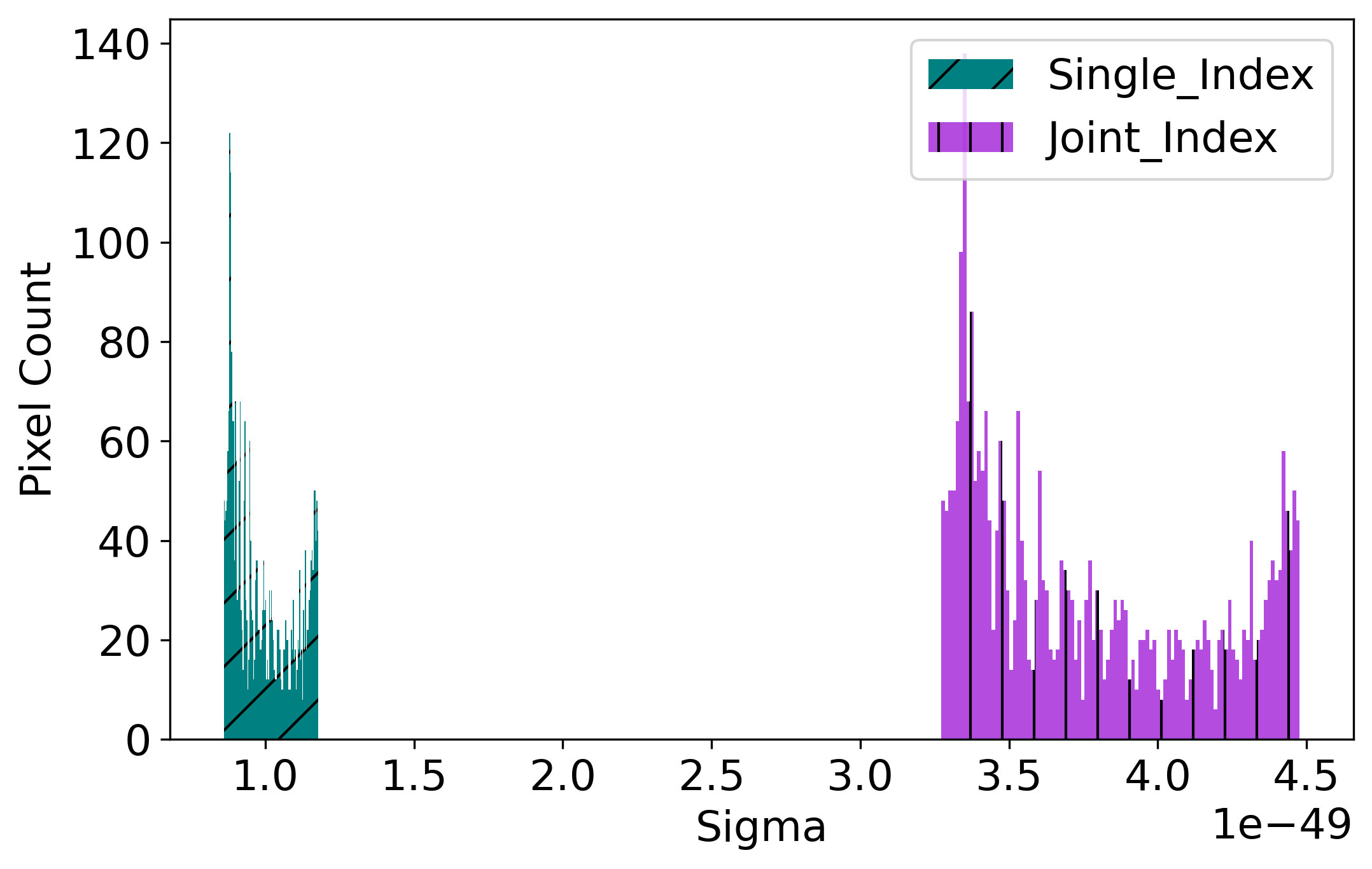}
    \includegraphics[width = 0.32\textwidth]{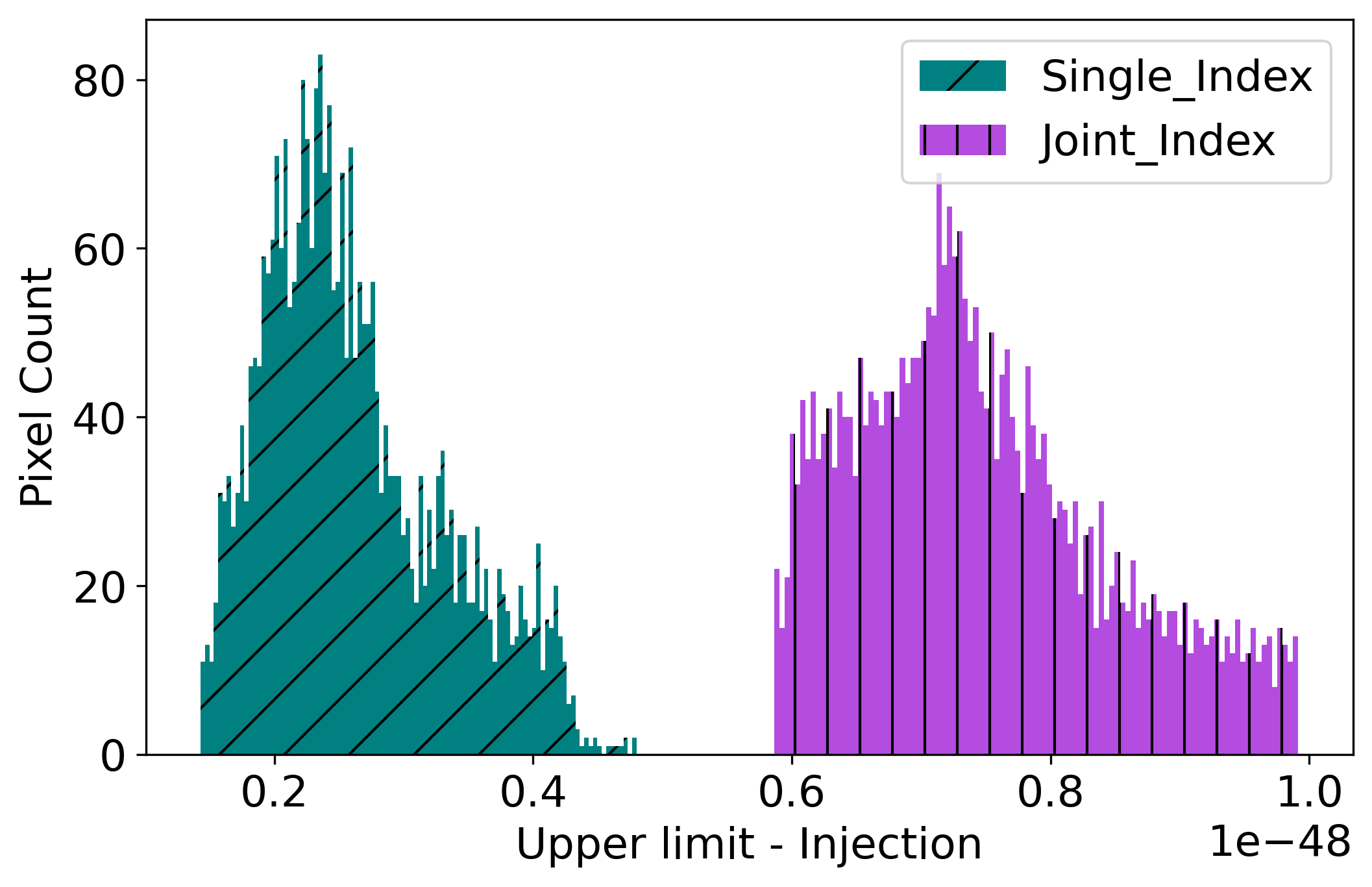}\\
    \includegraphics[width = 0.32\textwidth]{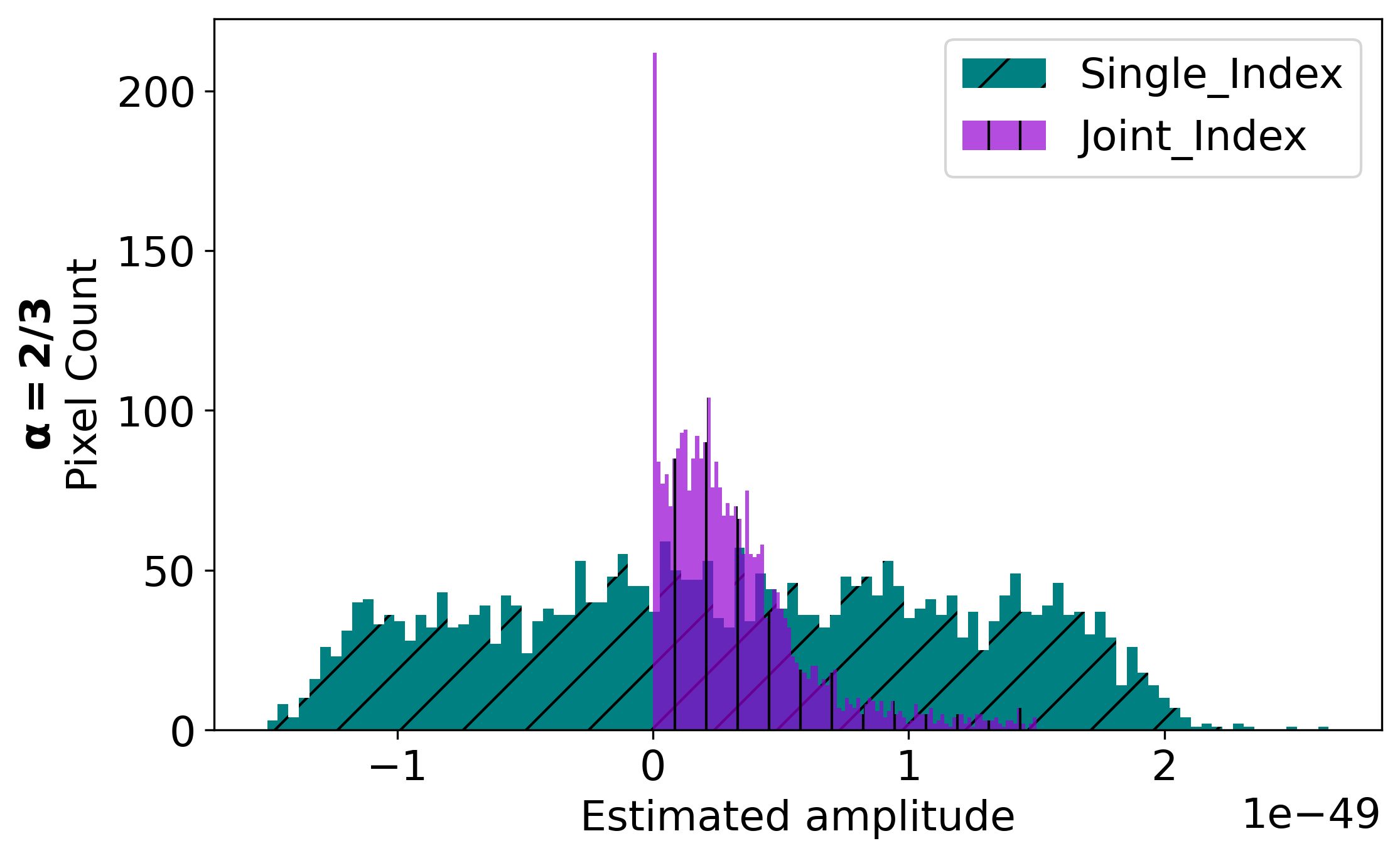}
    \includegraphics[width = 0.32\textwidth]{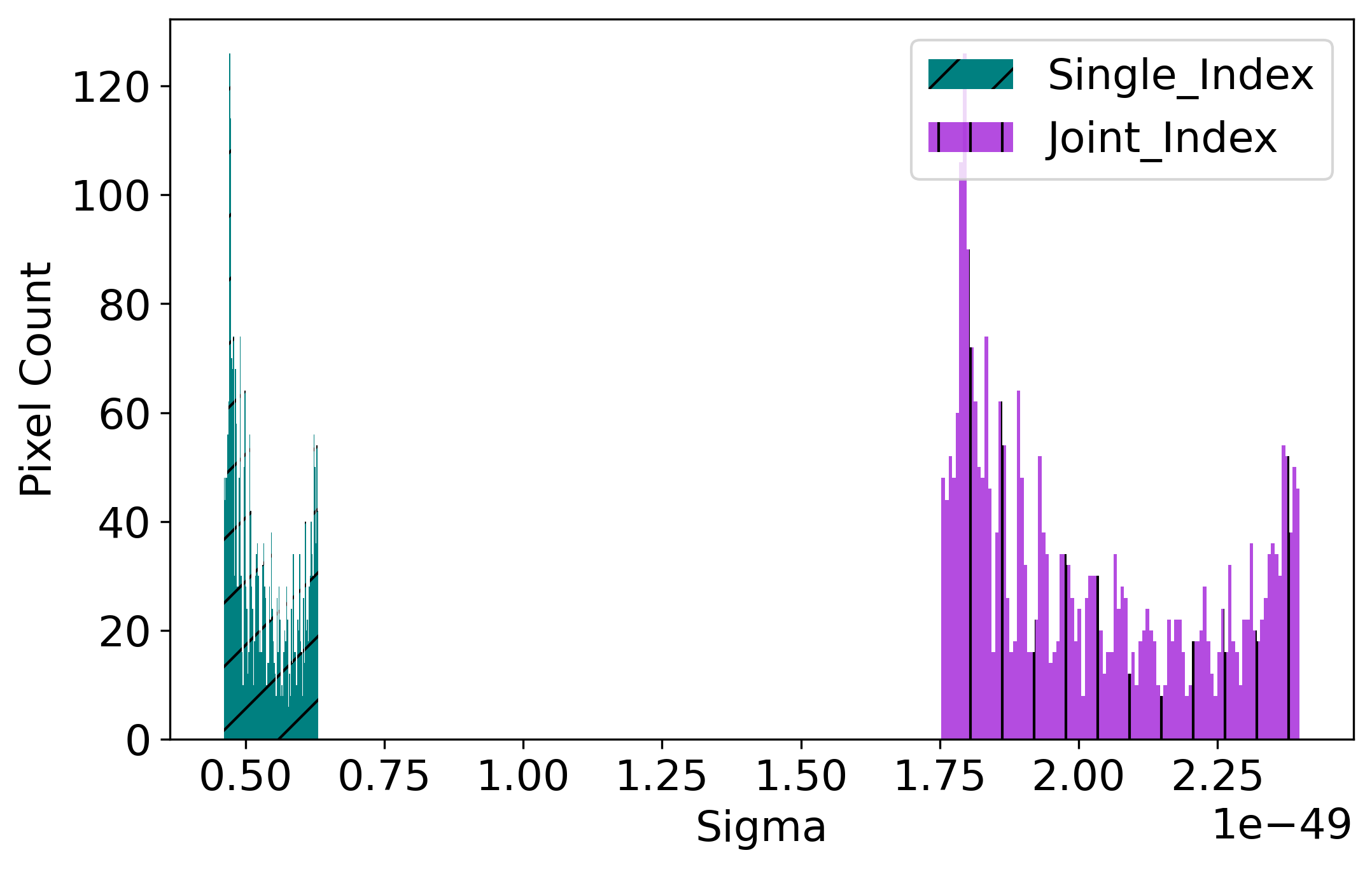}
    \includegraphics[width = 0.32\textwidth]{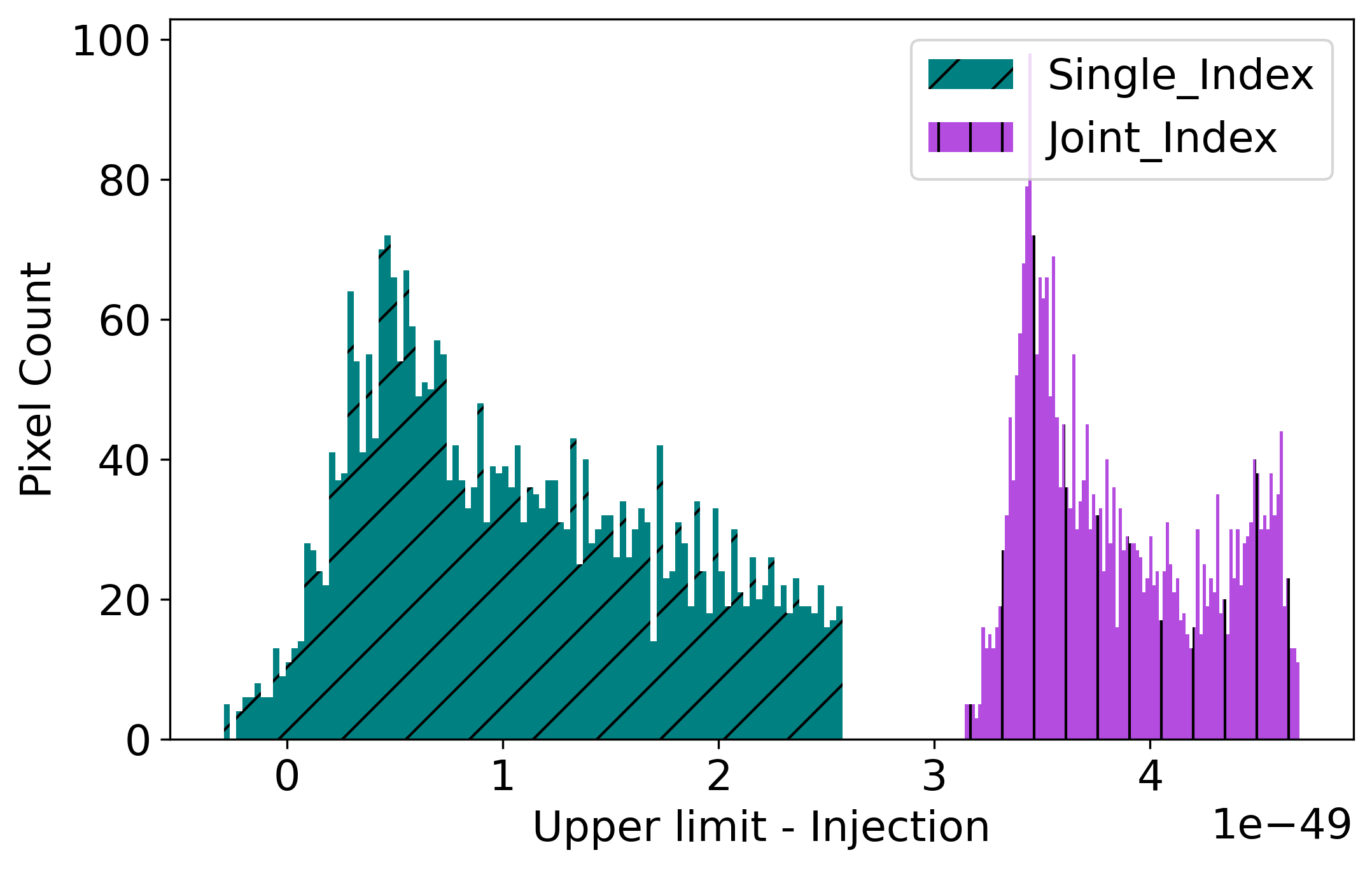}\\
    \includegraphics[width = 0.32\textwidth]{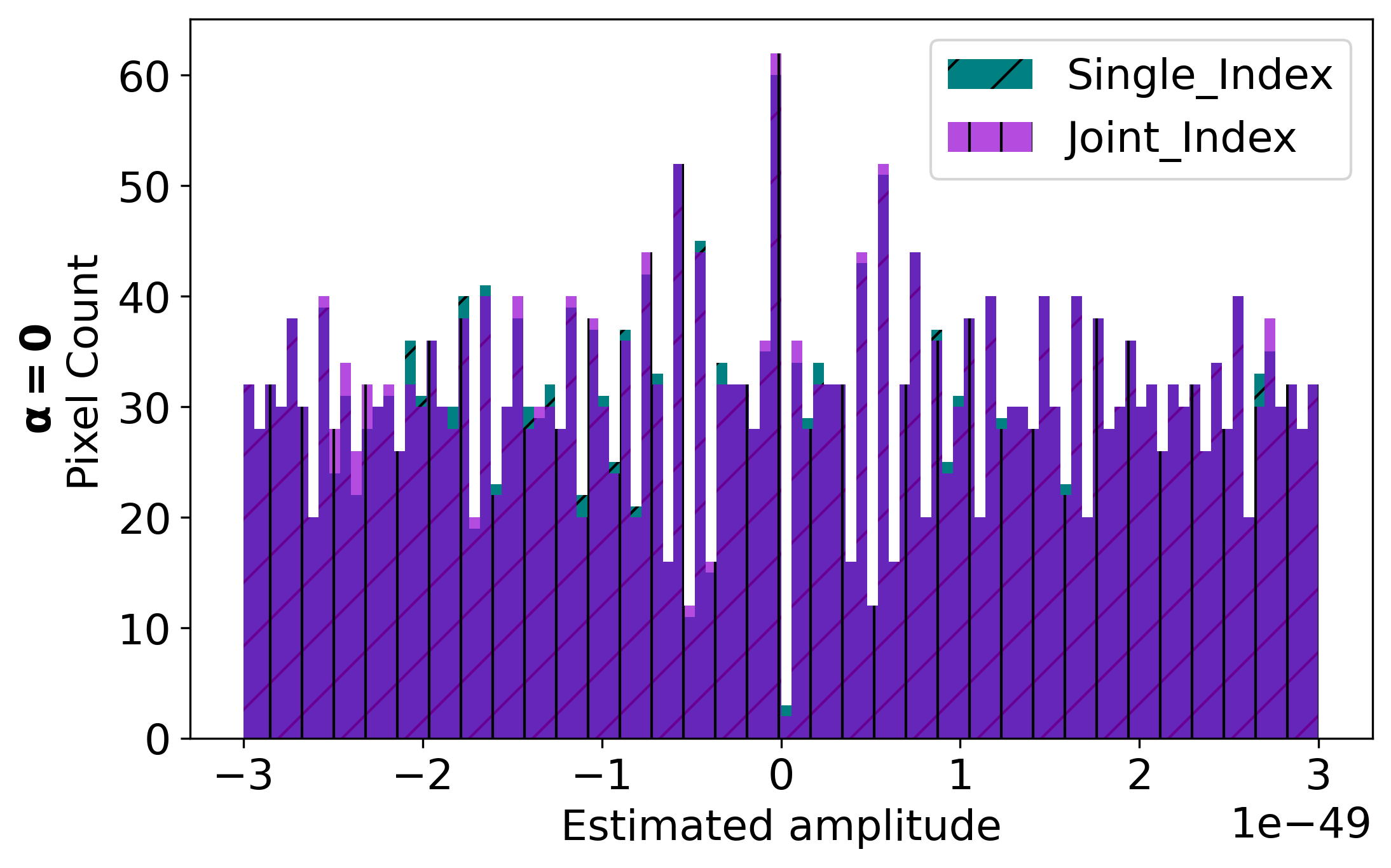}
    \includegraphics[width = 0.32\textwidth]{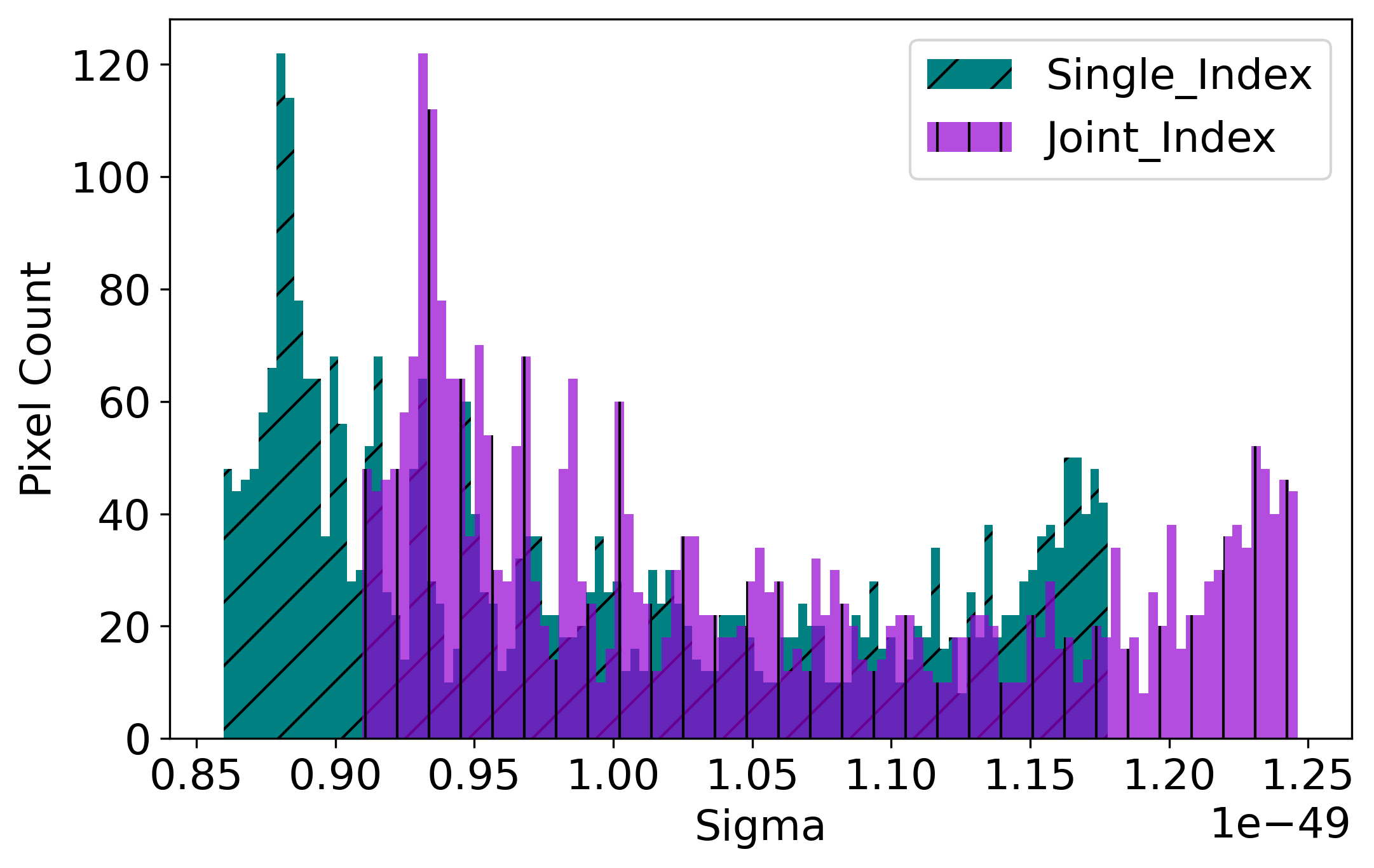}
    \includegraphics[width = 0.32\textwidth]{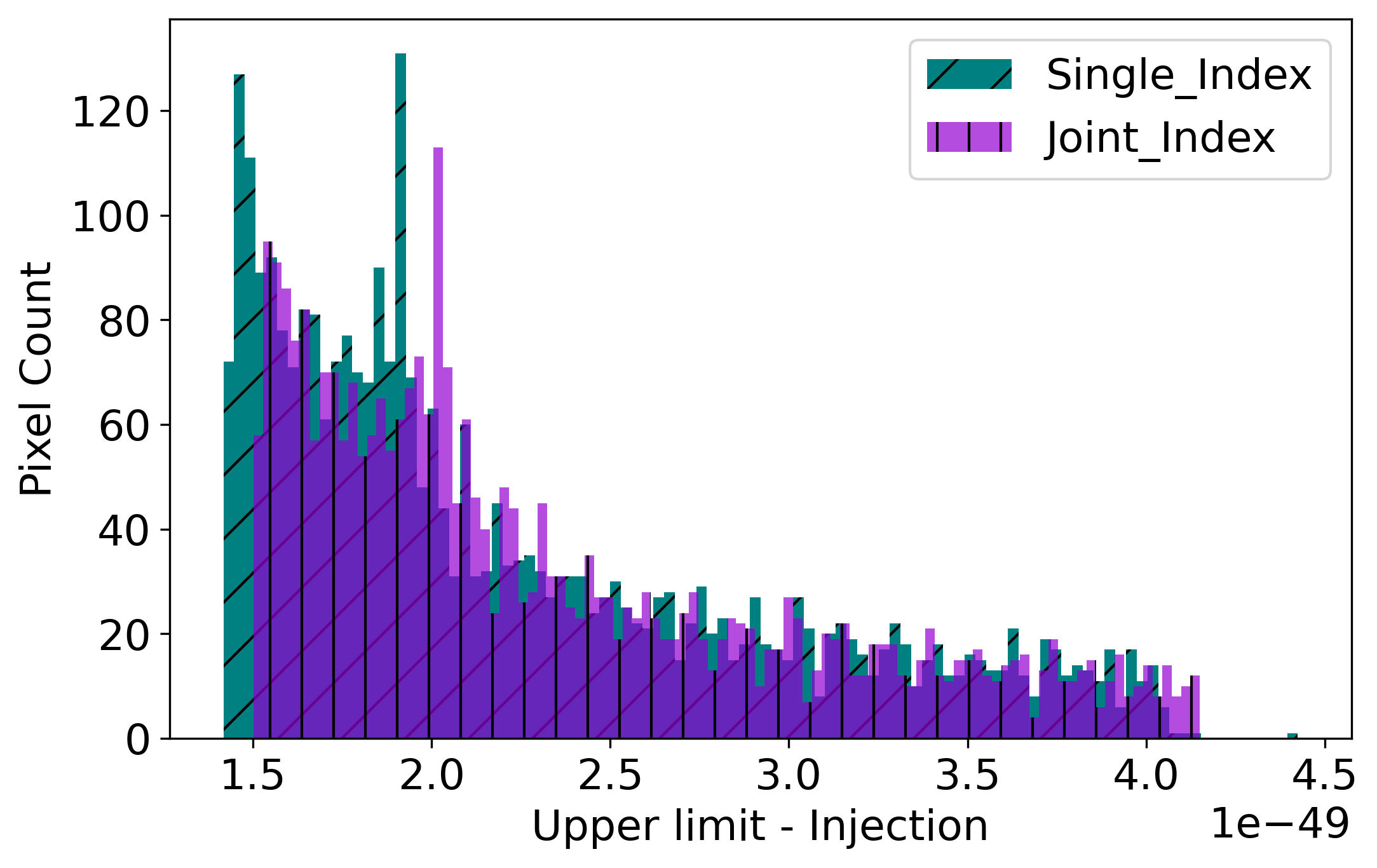}\\
    \includegraphics[width = 0.32\textwidth]{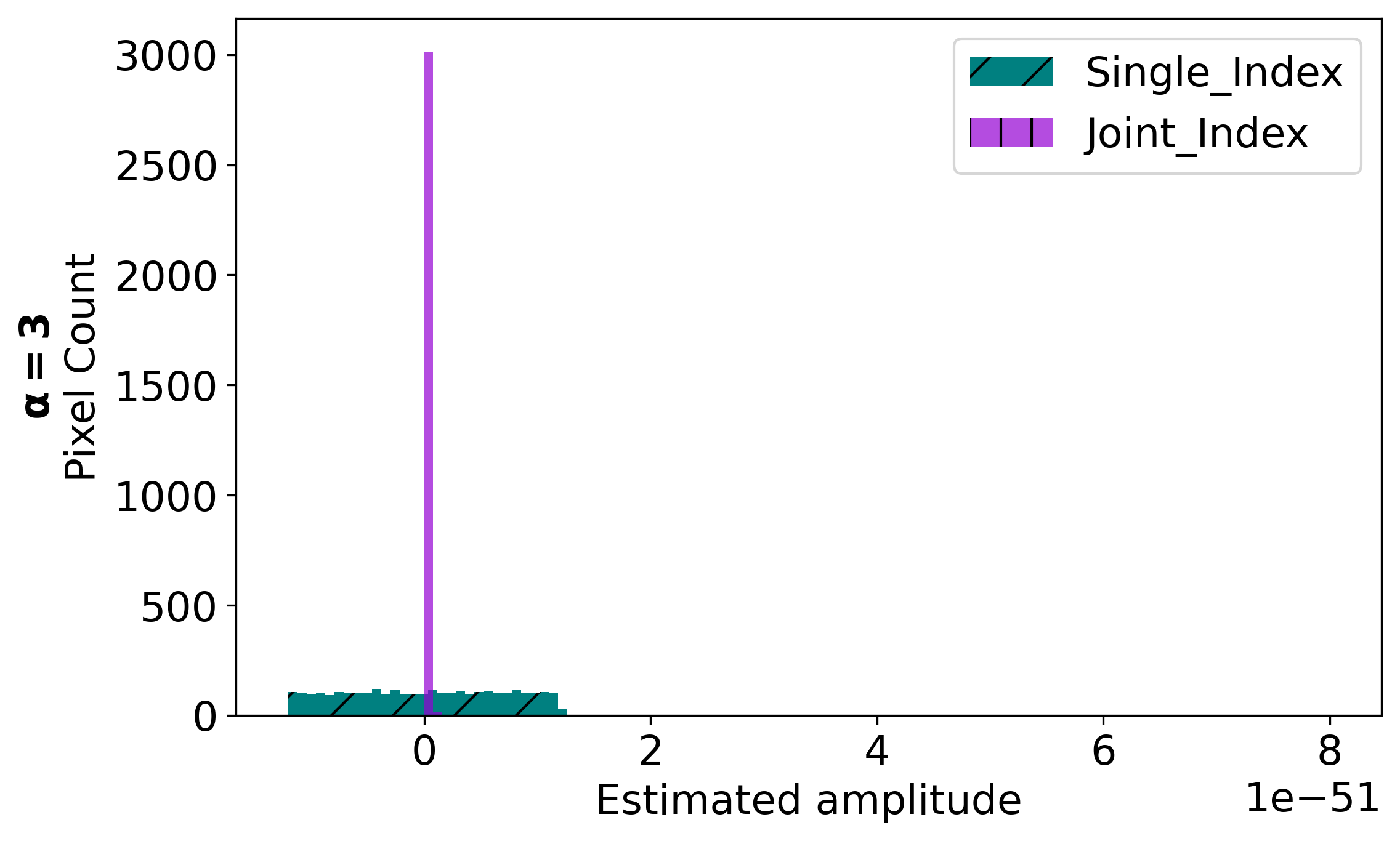}
    \includegraphics[width = 0.32\textwidth]{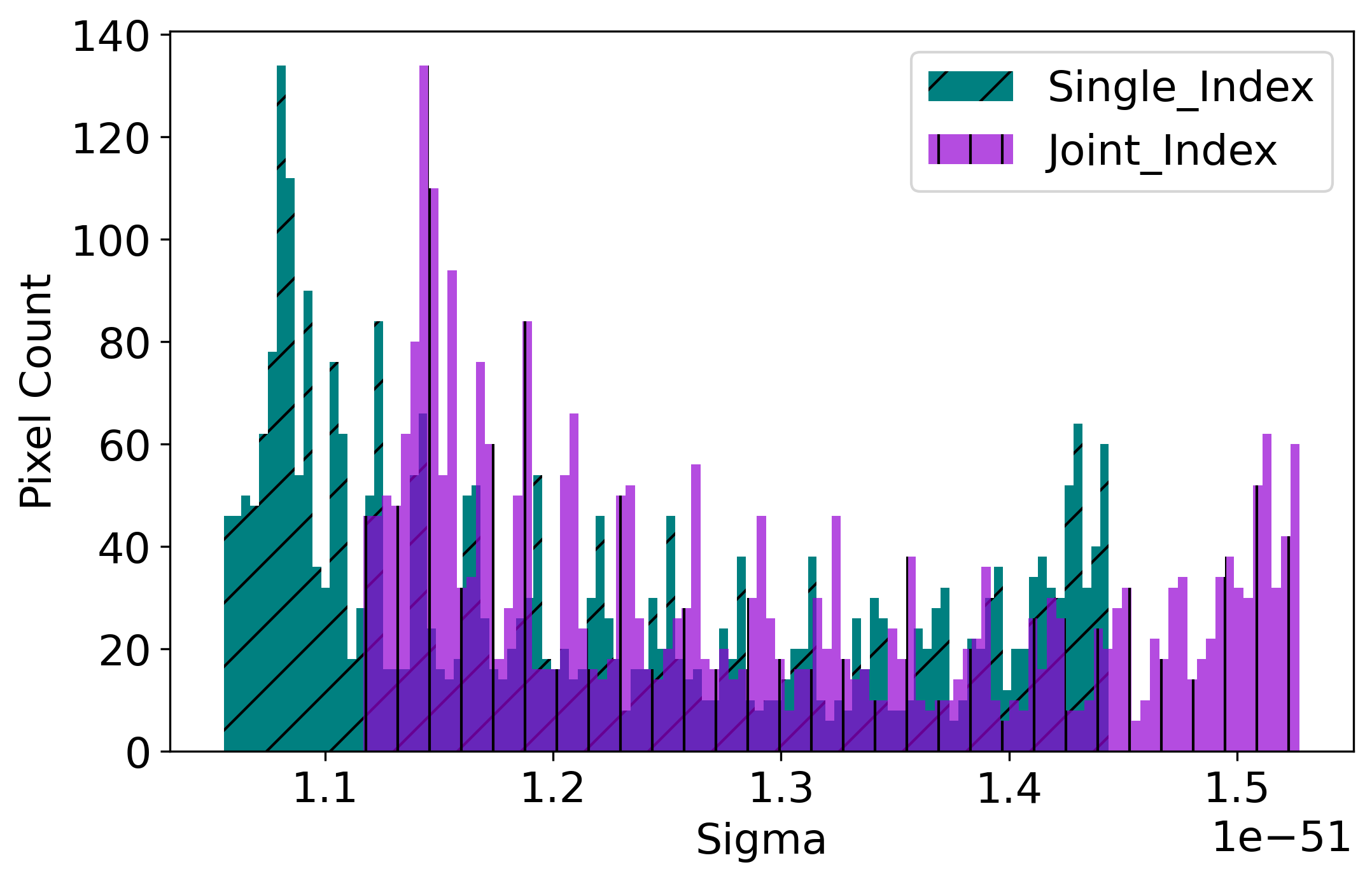}
    \includegraphics[width = 0.32\textwidth]{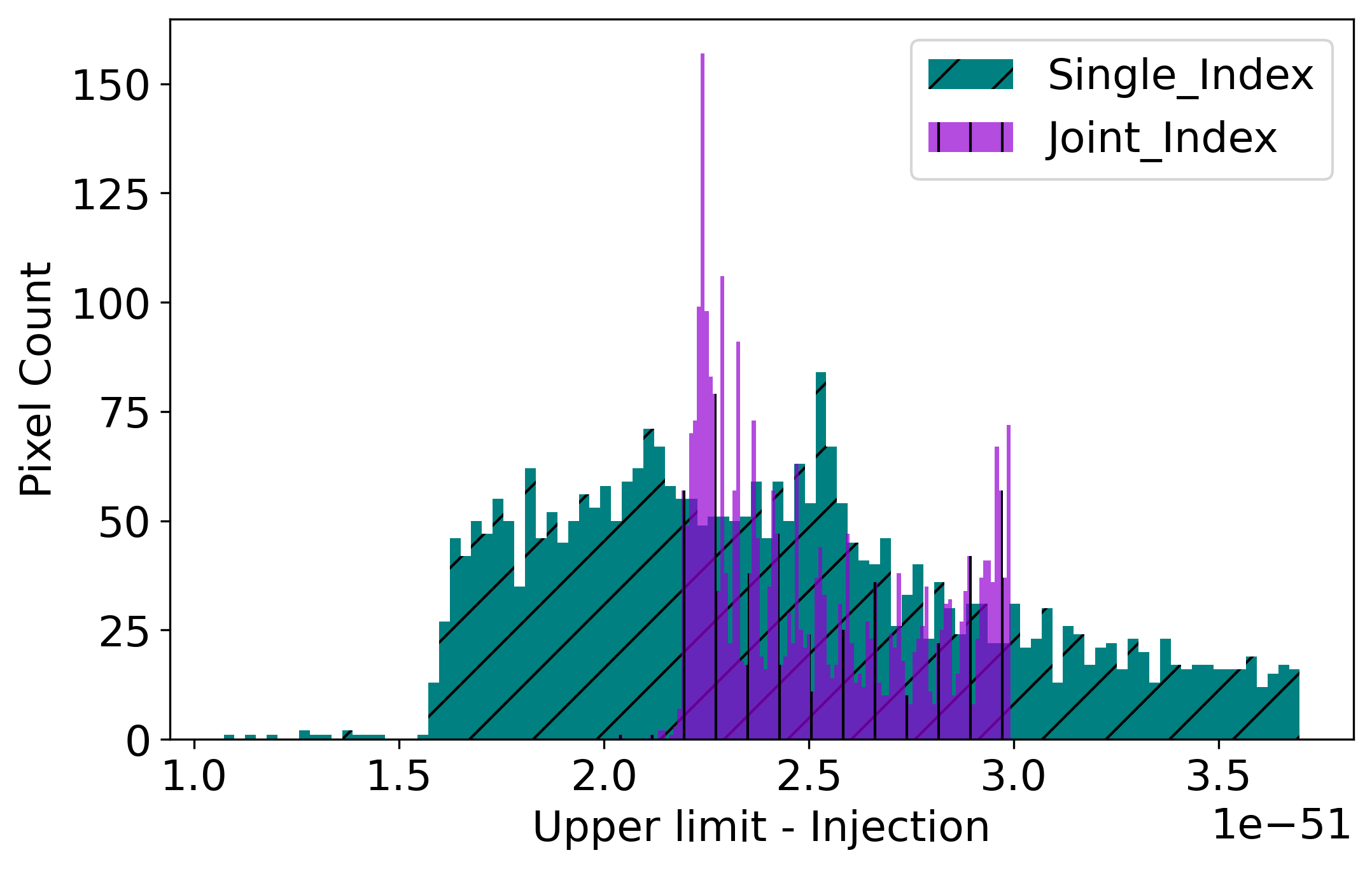} \\
    \includegraphics[width = 0.32\textwidth]{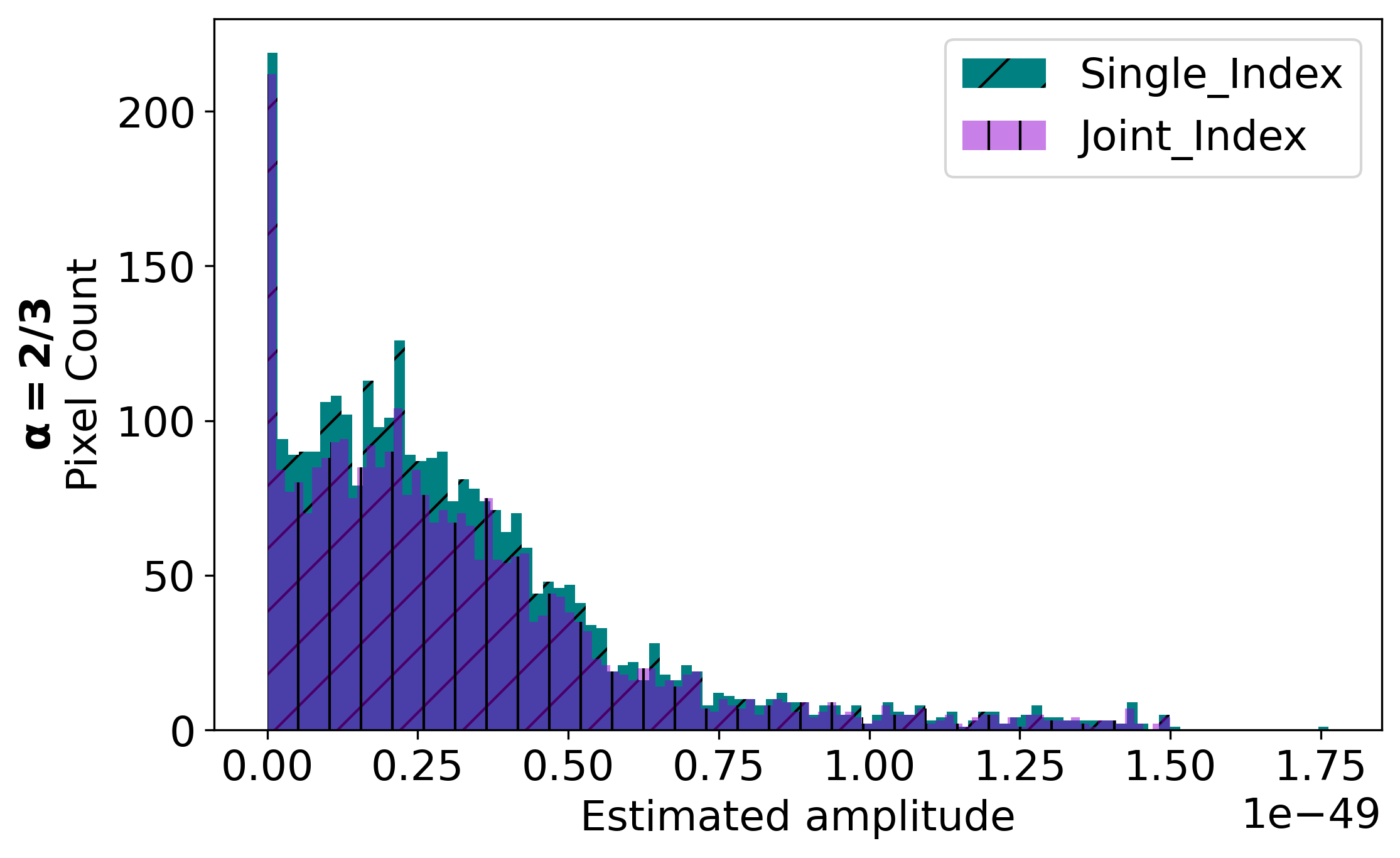}
    \includegraphics[width = 0.32\textwidth]{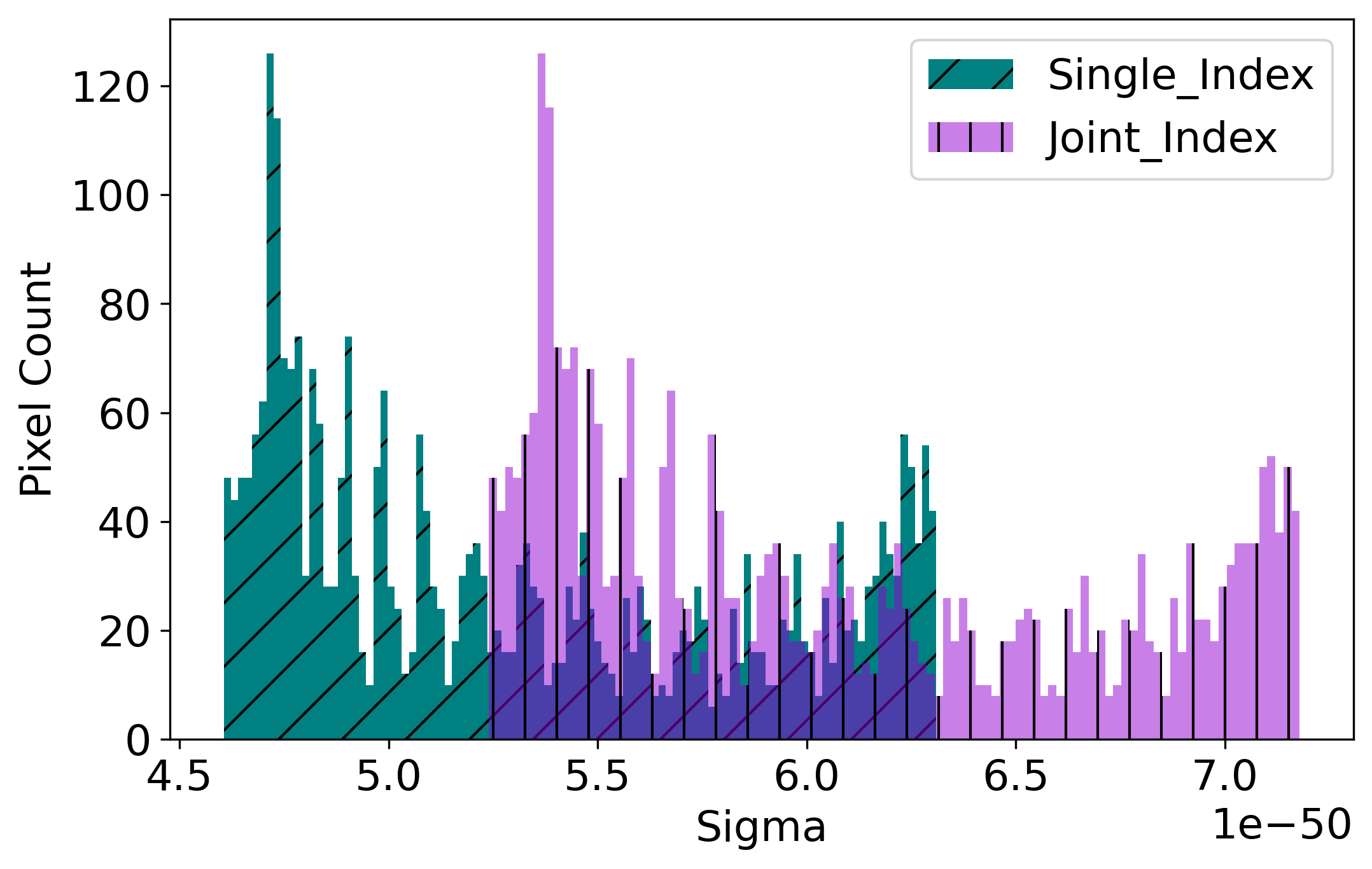}
    \includegraphics[width = 0.32\textwidth]{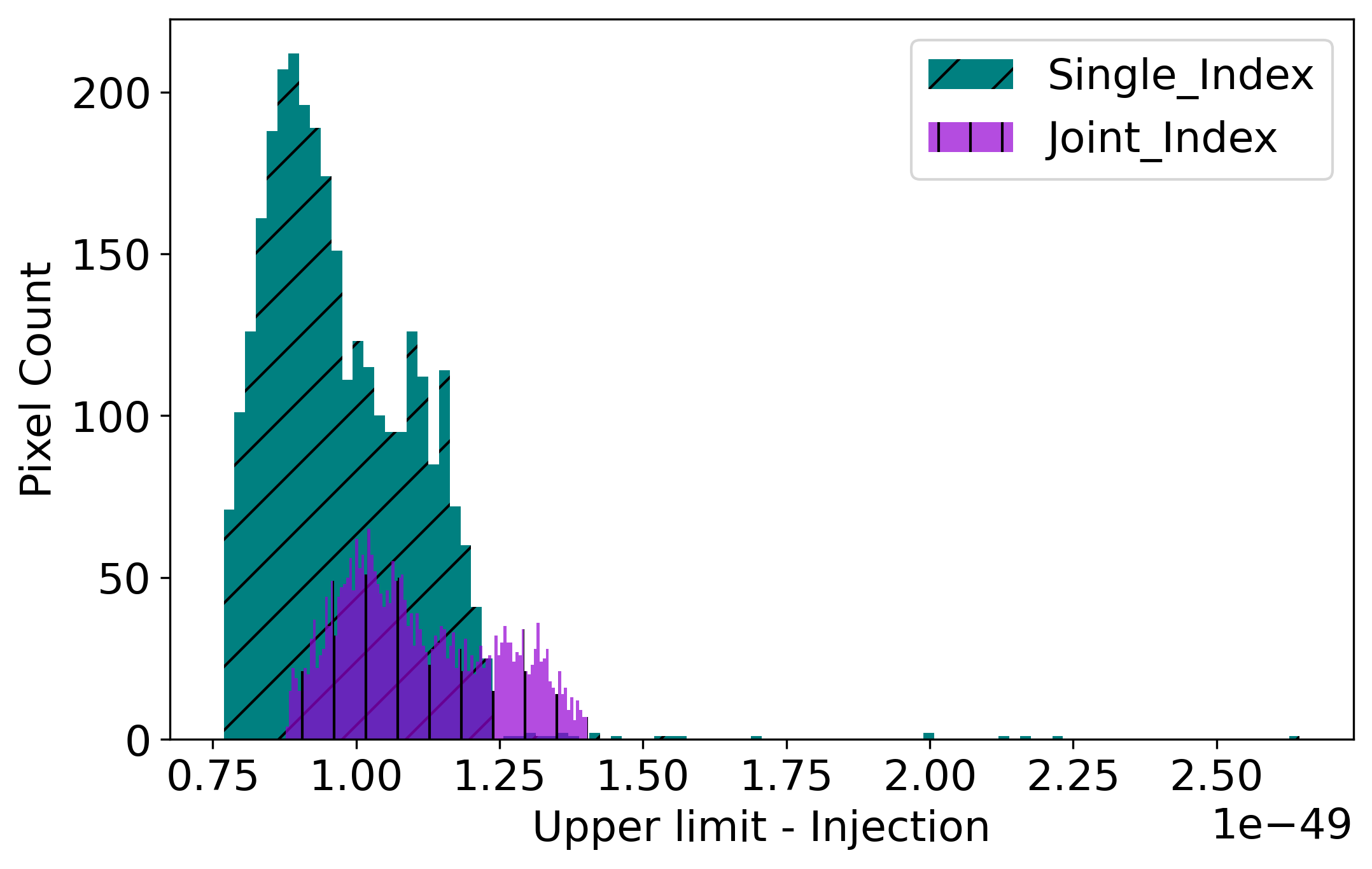}\\
    \includegraphics[width = 0.32\textwidth]{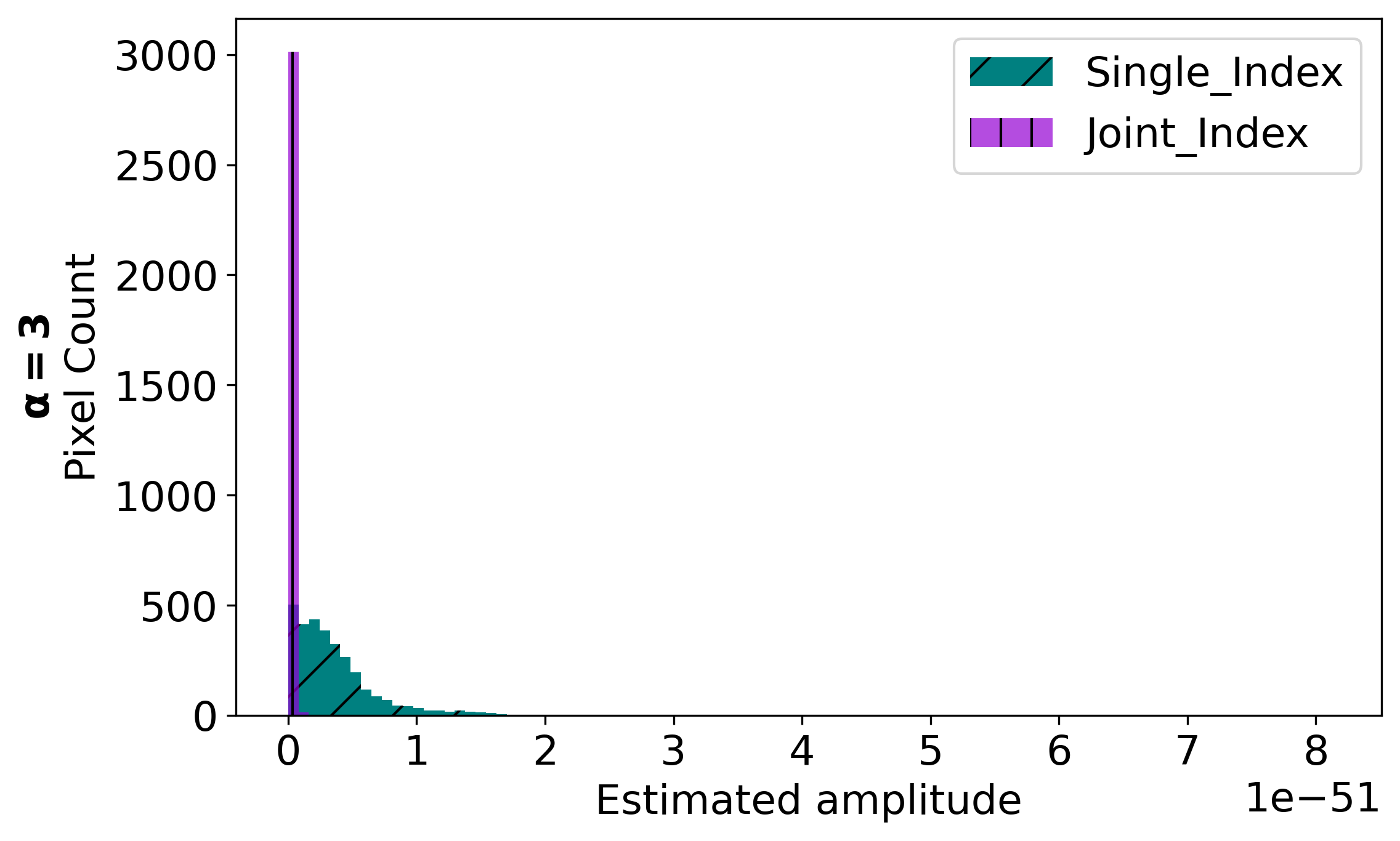}
    \includegraphics[width = 0.32\textwidth]{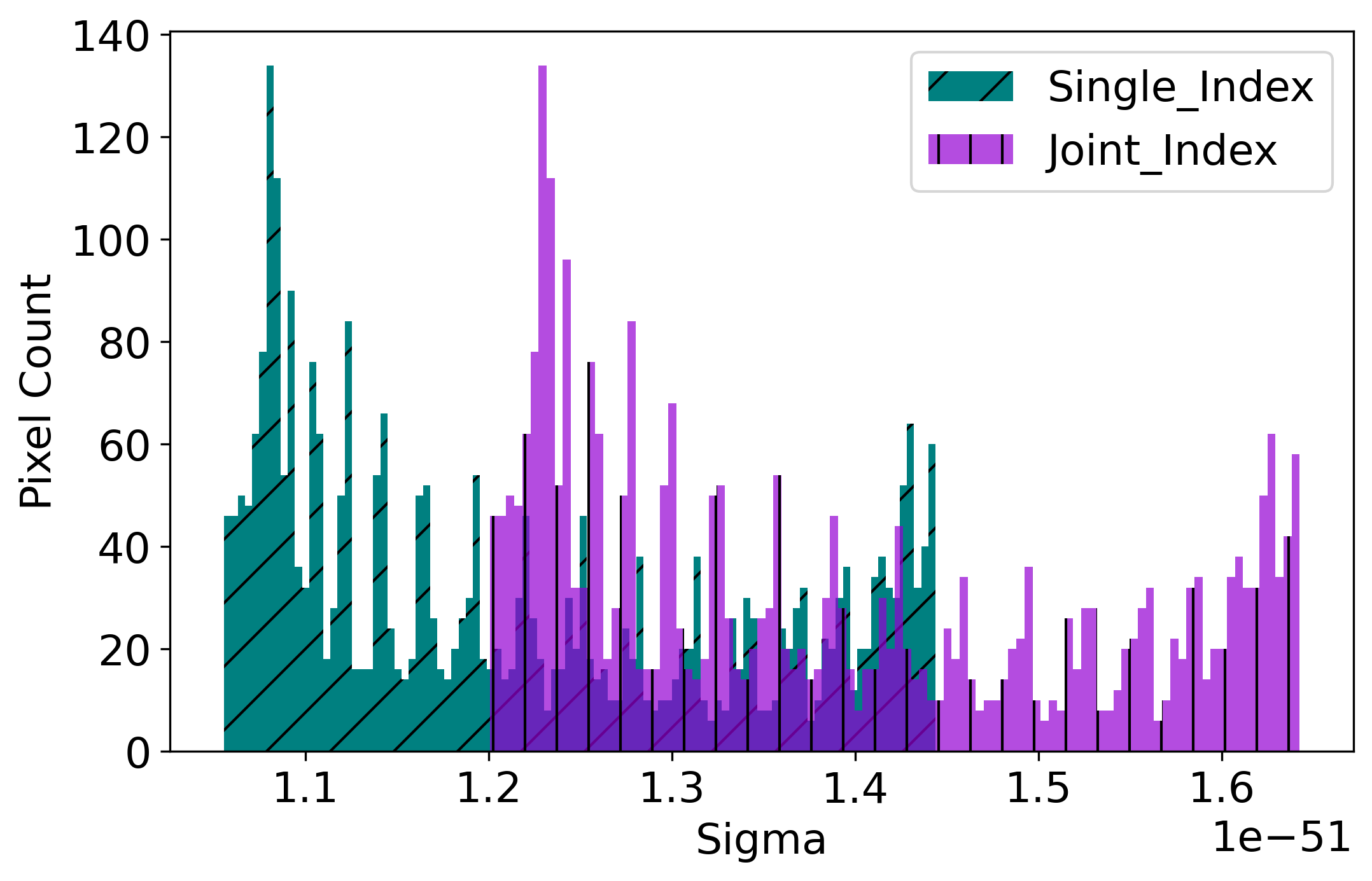}
    \includegraphics[width = 0.32\textwidth]{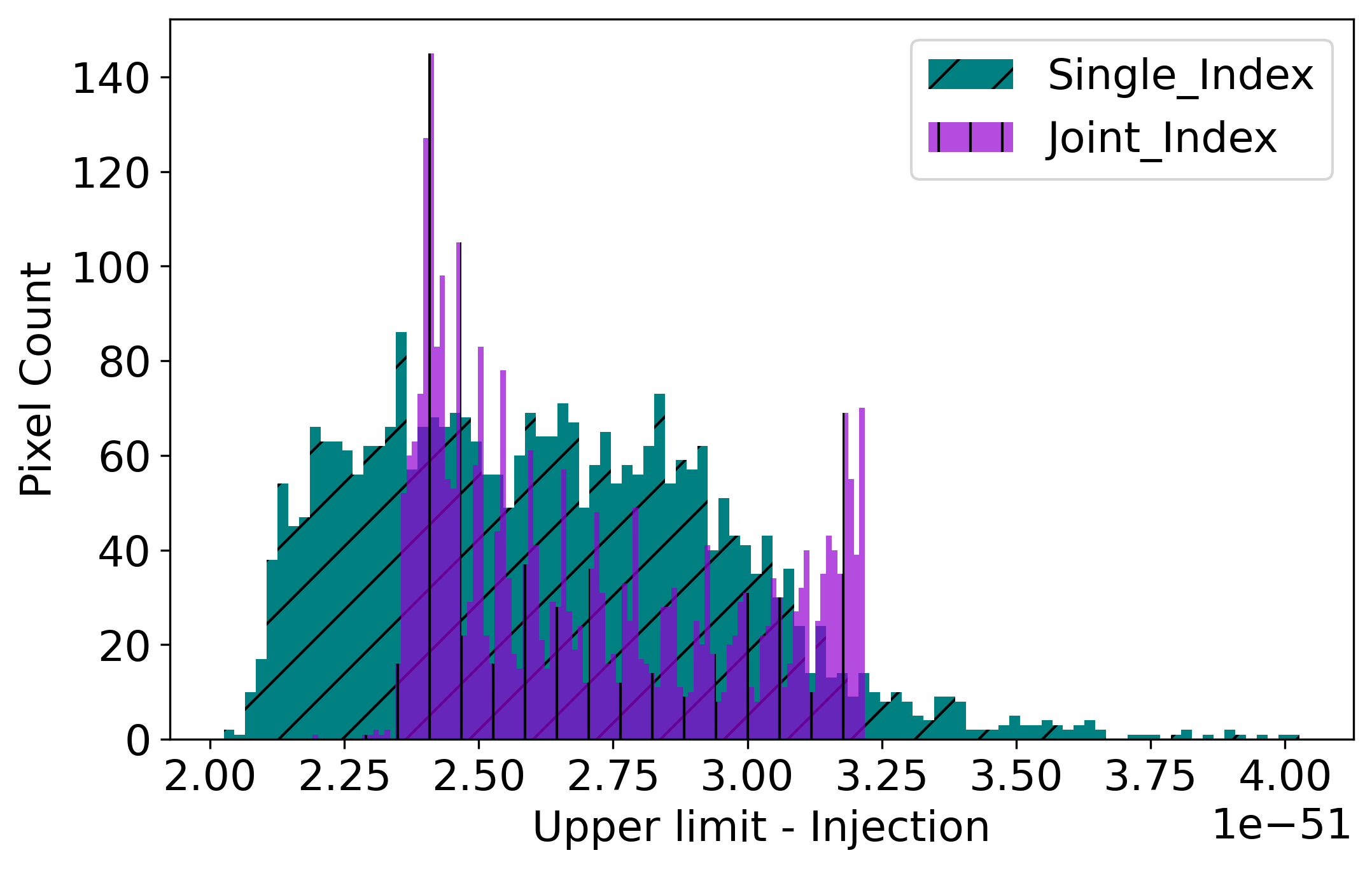}
    \caption{Estimated amplitude, variance and the difference between upper limit and injected strength (with corresponding injection amplitudes of $5\times10^{-49}, 2\times10^{-49}$ and  $8\times10^{-51}$ for $\alpha=(0,2/3,3)$ respectively) obtained from two-component analysis using both single-index (cyan slanted hatch) and joint-index (purple vertical hatch) analysis are shown respectively from left to right. Starting from top, each panel represent the results considering $\alpha=(0,2/3)$, $\alpha=(0,3)$ and $\alpha=(2/3,3)$ components.}
    \label{fig:2comp}
\end{figure*}

\begin{figure}
    \centering
    \includegraphics[width=\columnwidth]{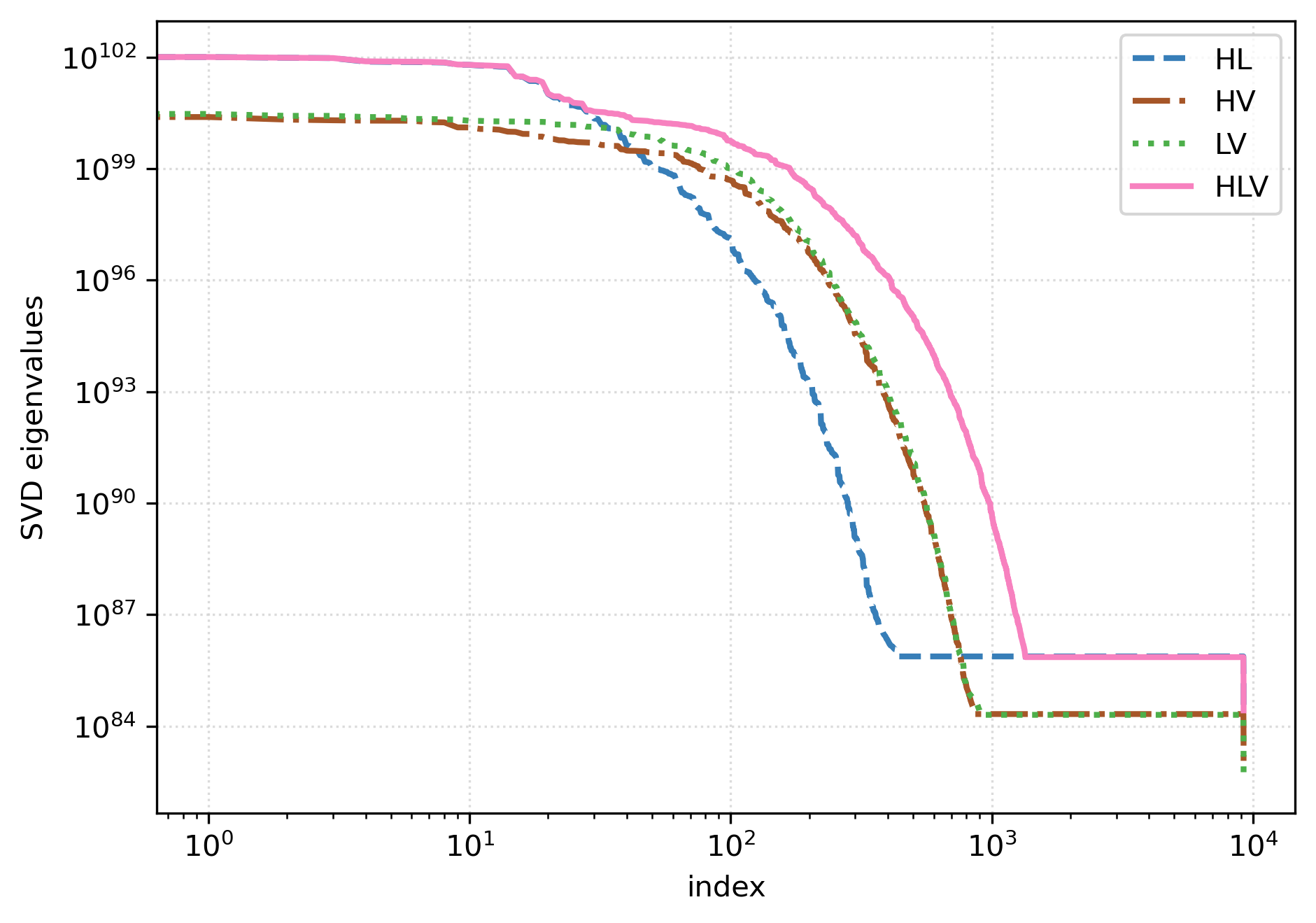}
    \caption{SVD eigenvalues of the coupling matrix $\vb{C}$ (considering both pixel-to-pixel and spectral covariance) for multiple baseline (H-Hanford, L-Livingston, V-Virgo). Solid line (pink) represents the results from combined baseline (HLV). It is evident that most dominant contributions to the eigenvalues are coming from the HL baseline. We perform this exercise using a \hpx resolution of $N_{\mbox{side}} = 8$.}
    \label{fig:multi_baseline}
\end{figure}

Increasing the number of components increases the uncertainty or variance associated with each component estimations. It has been shown that~\cite{Parida_isotropic} if we neglect the covariance between spectral shapes, we underestimate the error bar. On the other hand, the error bar increases when we simultaneously estimate the individual spectral shapes. We have shown this behavior in Figure~\ref{fig:2comp}, where we used the joint-index analysis by considering only two-component simultaneously. By comparing the histogram of the variance with the one for three-component case (see Figure~\ref{fig:ul_hist}) we can see that the variance reduces for all the two-component cases (it is also worth noting that, here also, the single-index analysis for $\alpha=2/3$, many points in the upper limit histogram (second panel from top)  fall under the negative region). This suggests that there is a trade-off between accurately extracting the individual components and minimizing the uncertainties. In general, the addition of detectors to the network improves this trade-off problem. However, given the current sensitivity of Virgo detector~\cite{virgo}, adding it to our analysis will not contribute considerably. To demonstrate these effects, we consider all the three baselines (HL, HV, LV) and eigenvalues of their respective coupling matrix (here we consider both pixel-to-pixel and spectral index covariance). Eigenvalues obtained from the singular value decomposition are plotted and compared with the combined baseline result (HLV) in Figure~\ref{fig:multi_baseline}. It is evident from this figure that, given the sensitivity of the Virgo detector, the most dominant contributions to the eigenvalues are coming from the HL baseline. This situation can improve drastically when detectors like KAGRA~\cite{Kagra,kagra_obsScience} and LIGO-India~\cite{ligo_india,LIGO_india_obsScience} start their observation run. 

\bibliography{biblio}

\end{document}